\documentclass{aa}

\usepackage{graphicx}
\usepackage{natbib}

\usepackage{amsmath}    
\usepackage{amssymb}    

\begin{document}

\title{New analytical solutions for chemical evolution models: characterizing the population of star-forming and passive galaxies}

\author {E. Spitoni\inst{1} \thanks {email to:
    spitoni@oats.inaf.it} \and F. Vincenzo\inst{1,2} \and F. Matteucci\inst{1,2}} \institute{
  Dipartimento di Fisica, Sezione di Astronomia,  Universit\`a di Trieste, via
  G.B. Tiepolo 11, I-34131, Trieste, Italy \and I.N.A.F. Osservatorio
  Astronomico di Trieste, via G.B. Tiepolo 11, I-34131, Trieste,
  Italy}

\date{Received xxxx / Accepted xxxx}

\abstract
{Analytical models of chemical evolution, including inflow and outflow
of gas, are important tools for studying how the metal content in galaxies
evolves as a function of time.}{ We present new
analytical solutions for the evolution of the gas mass, total mass, and
metallicity of a galactic system when a decaying exponential infall
rate of gas and galactic winds are assumed. We apply our model to
characterize a sample of local star-forming and passive galaxies from
the Sloan Digital Sky Survey data, with the aim of reproducing their
observed mass-metallicity relation.}{We derived how the two populations
of star-forming and passive galaxies differ in their particular
distribution of ages, formation timescales, infall masses, and mass
loading factors.}  {We find that the local passive galaxies are, on
average, older and assembled on shorter typical timescales than the
local star-forming galaxies; on the other hand, the
star-forming galaxies with higher masses generally show
older ages and longer typical
formation timescales compared than star-forming
galaxies with lower masses. The local star-forming galaxies experience stronger galactic
winds than the passive galaxy population. Exploring the effect of
assuming different initial mass functions in our model, we show that to reproduce the observed mass-metallicity relation, stronger
winds are requested if the initial mass function is top-heavy.
Finally, our analytical models predict the assumed sample of local
galaxies to lie on a tight surface in the 3D space defined by stellar
metallicity, star formation rate, and stellar mass, in agreement with the
well-known fundamental relation from adopting gas-phase
metallicity.}{By using a new analytical model of
chemical evolution, we characterize an ensemble of SDSS galaxies in
terms of their infall timescales, infall masses, and mass loading
factors.  Local passive galaxies are, on average, older and
assembled on shorter typical timescales than the local star-forming
galaxies. Moreover, the local star-forming galaxies show stronger galactic
winds than the passive galaxy population.  Finally, we find that the
fundamental relation between metallicity, mass, and star formation rate for these local galaxies is still valid when
adopting the average galaxy stellar metallicity. }

\keywords{galaxies: abundances - galaxies: evolution - galaxies: ISM}

\titlerunning{Chemical evolution of  SF and passive galaxies  }
\authorrunning{Spitoni et al.}
\maketitle

\section{Introduction}

 Chemical evolution of galaxies studies how subsequent stellar
 generations and gas flows alter the chemical composition of the
 galaxy interstellar medium (ISM) to give rise to the present-day
 observed chemical abundance pattern of galaxies.  In this respect, the
 galaxy star formation and gas mass assembly histories play a major
 role, together with the assumed initial mass function (IMF) and
 stellar nucleosynthetic yields.  Chemical evolution is an essential
 ingredient in the broader framework of galactic archeology, which aims at
 constraining and recovering the formation and evolution of galaxies,
 starting from the observed chemical, dynamical, and photometric
 properties of their member stars at the present time.

 Simple analytical models of chemical evolution have previously
enabled deriving analytical functions for the metallicity evolution of
 a stellar system.  We assume a  constant IMF and complete mixing of the various chemical
 species within the ISM of a galaxy at any time of its evolution.
 Finally, a further fundamental hypothesis is to retain the instantaneous recycling approximation  (IRA): all the
 stars with mass $m\ge1\,\mathrm{M}_{\sun}$ instantaneously die as
 they form,  whereas all
 the stars with $m<1\,\mathrm{M}_{\sun}$ have infinite lifetimes. This
 type of models still represents a useful tool for tracing the metallicity
 evolution of galaxies, but only when the abundance of chemical
 elements produced on typical short timescales is considered. An
 example of such a chemical element is given by oxygen, which
 incidentally represents the best proxy for the total metallicity of
 the galaxy ISM.  Pioneering works in this field are considered those
 of \citet{schmidt1963}, \cite{searle1972}, \citet{tinsley1974}, and \citet{pagel1975}.

To build a
realistic analytical model for the galaxy chemical evolution, gas
flows must be included in the set of differential equations to be
solved because galaxies do not evolve as closed boxes.  Analytical  and semi-analytical solutions for models including infall and outflow
of gas have been known for at least 30 years (see, for
example,  \citealt{chiosi1980,clayton1986, clayton1993,pagel1975,hartwick1976,clayton1988,twarog1980,edmunds1990,erb2008,recchi2015,kudritzki2015}).
Moreover, solutions have also been found later with radial gas flows,
provided a particular gas velocity profile is assumed in the equations
(see \citealt{lacey1985, edmunds1995,martinelli1998,portinari2000,spitoni2011,pezzulli2015}).  A summary of some of the
most frequently used analytical solutions for the metallicity evolution of a
system with different prescriptions for inflows and outflows of gas
can be found in \citet{recchi2008} and \citet{spitoni2010}.

Recently, \citet{spitoni2015} presented an analytical solution for the
evolution of the metallicity of a galaxy in presence of environmental
effects.  In this
work, a galaxy suffers  the infall of enriched
gas from another evolving galactic system, with the
metallicity of the infalling material evolving according to the
metallicity evolution of the companion galaxy, hence with chemical abundances variable in time.

More recently, \citet{weinberg2016} showed chemical analytical solutions with different prescriptions for the star formation
rate (SFR) analyzing constant, exponential, or linear-exponential star formation histories.

 To solve the set of differential equations for the gas mass,
 total mass, and metallicity of the galaxy and hence obtain analytical
 solutions for these quantities, some of the previous works in the
 literature (see, for
 example, \citealt{matteucci1983}, \citealt{matteucci2012}) assumed an
 infall rate of gas that is directly proportional to the SFR over the entire galaxy evolution.  This
 assumption is not physical, and it represents a strong simplification
 in analytical models. Nevertheless, by exploring the effects of
 different prescriptions for the infall term in the equations, the
 final predicted physical properties of the galaxy have been found to not
deviate substantially from the case of a generic exponential
 infall law \citep{recchi2008}.   The infall of gas
 that follows an  exponential law is a fundamental assumption
 adopted in most of the detailed numerical chemical
 evolution  models  in which  IRA is relaxed.  Chemical evolution models
 of our Galaxy (\citealt{chiappini1997,romano2010,brusadin2013,
 micali2013}) assume that the various different stellar components
 formed through different separated accretion episodes of gas, with
 the  accretion rate of each episode obeying to a decaying
 exponential law.  \citet{colavitti2008} reported that the two-infall
  model of \citet{chiappini1997} is qualitatively in
 agreement with results of   the GADGET2
 (\citealt{springel2005}) cosmological hydrodynamical
 simulations   when the standard cosmological
 parameters from WMAP three-years (\citealt{spergel2007}) are
assumed, namely
 $\Omega_0$=0.275, $\Omega_{\lambda}$=0.725, and $\Omega_b$=0.041.

We here present the results of an analytical chemical
evolution model in which a decaying exponential infall rate of gas is
assumed as a function of time; we show that analytical solutions for
the evolution of the galaxy metallicity, gas mass, and total mass can
be found under this assumption.  Furthermore, we apply our model to
investigate and explain the observed mass-metallicity relation (MZ
relation, hereafter), as derived in a sample of galaxies from the
Sloan Digital Sky Survey (SDSS) by \citet{peng2015}. We
extend and update the methods and results obtained
by \citet{spitoni2010}, which reproduced the observed MZ relation in
$27730$ local SDSS star-forming galaxies \citep{kewley2008} with an
analytical model of chemical evolution.  In particular, we aim at
characterizing the two distinct MZ relations that \citet{peng2015}
derived for the local actively star-forming (we refer to them
as \textit{star-forming}) galaxies and the passively evolving
(\textit{passive}) galaxies.  A tight relation between the stellar
mass and the gas-phase metallicity such as displayed by a large sample of
galaxies like the one provided by the SDSS can be used to constrain
the various fundamental parameters playing a role in any galaxy
formation theory: the  infall timescale,  wind loading factor
parameters, infall mass values, and the star formation efficiencies.

On the one hand, the analytical model of \citet{spitoni2010} assumed
both the gas outflow rate and the gas infall rate to be directly
proportional to the galaxy SFR; when this simplifying
assumption is made, the analytical solution for the
metallicity of the system does not explicitly depend upon the time
variable, which indeed turns out to be hidden in the galaxy gas mass
fraction entering in the equations. 

 Furthermore,  \citet{kudritzki2015}  assumed constant ratios of galactic
wind mass-loss and accretion mass gain to SFR in the IRA approximation. They investigated  the radially averaged metallicity distribution
of the interstellar medium   of a
sample of 20 local star-forming disk galaxies   by means of  analytical
chemical evolution model.

On the
other hand, in this work, we show that when  a
decaying exponential infall law is assumed, the analytical solution for the
evolution of the galaxy metallicity explicitly depends upon the time
variable; this fact will allow us to also provide an estimate for the age
of the galaxies and characterize them in terms of their infall
timescale.

\citet{mannucci2010} added a further dimension to the observed 
 MZ relation of local galaxies; in particular, they found that local
galaxies place themselves on a tight surface in the 3D space defined
by gas phase metallicity, SFR, and stellar mass, with a
small residual dispersion of about $0.05\,\mathrm{dex}$ and hence
showing an underlying fundamental relation.  In this work, we test
whether our new analytical model is able to recover a similar
fundamental relation for the local population of passive and star-forming
galaxies.

Our paper is organized as follows.  In Sect. 2, after summarizing
the basic assumptions of our chemical evolution model, we present the
new analytical solutions for the galaxy metallicity, gas mass, and
total mass.  In Sect. 3 we discuss the observed MZ relation in the
population of local passive and star-forming galaxies by Peng et
al. (2015) and present the method we employ to reproduce it.  In
Sect. 4 we present the methods used to reproduce the MZ relation
with the new analytical solutions. In Sect. 5 we report our results,
and in Sect. 6, we discuss whether the population of galaxies drawn
by our new analytical model are predicted to follow the fundamental
relation of \citet{mannucci2010}. Finally, our conclusions are drawn
in Sect. 7.

 \begin{table*}
\begin{tabular}{c c c c c c c c c c }
\hline
\\
\multicolumn{2}{c}{}&
\multicolumn{4}{c}{\citet{salpeter1955}}&
\multicolumn{4}{c}{\citet{chabrier2003}}\\
\\
\multicolumn{2}{c}{}&
\multicolumn{2}{c}{{\it \normalsize Passive Galaxies}}&
\multicolumn{2}{c}{{\it \normalsize Star-Forming galaxies}}& 
\multicolumn{2}{c}{{\it \normalsize Passive Galaxies}}&
\multicolumn{2}{c}{{\it \normalsize Star-Forming galaxies}} \\
\\
\multicolumn{2}{c}{}&
\multicolumn{1}{c}{{\it \normalsize Range}}&
\multicolumn{1}{c}{{\it \normalsize $75\,\%$}} &
\multicolumn{1}{c}{{\it \normalsize Range}}&
\multicolumn{1}{c}{{\it \normalsize $75\,\%$}} &
\multicolumn{1}{c}{{\it \normalsize Range}}&
\multicolumn{1}{c}{{\it \normalsize $75\,\%$}} &
\multicolumn{1}{c}{{\it \normalsize Range}}&
\multicolumn{1}{c}{{\it \normalsize $75\,\%$}} 

 \\
\\

Infall timescale & $\tau$ [Gyr] & $0.07-4.8$ & $\le1.8$ & $0.07-7.8$& $\le6$ & $0.07-4.2$& $\le2.4$ & $0.07-7.8$ & $\le6.6$ \\
& &  &  &  &  \\

Age &  [Gyr] & $1.8-13.1$ & $\le9.6$ & $0-4.8$ & $\le2.4$ & $0.6-13.1$ & $\le10.2$ & $0-3$ & $\le1.8$ \\
& &  &  &  &  \\

Wind parameter & $\lambda$ & $0-1.5$ & $\le0.9$ & $0-5.8$ & $\le2.25$ & $0.6-3$ & $\le2.2$ & $0.6-10.2$ & $\le5.3$ \\
\\
Infall Mass & $\log(M_\mathrm{inf} [{\mbox M} _{\sun}])$  & $9.5-11.5$ & $\le10.5$ & $10.1-11.5$ & $\le11.43$ & $10.2-11.8$ & $\le10.9$ & $10.8-12.5$ & $\le12.05$ \\
 & &    \\

\hline
\end{tabular}
\caption{Principal characteristics of the computed passive and star-forming galaxies. In each column we indicate the range  spanned by the values  related to the infall timescale $\tau$,  the ages of the computed local  galaxies, the wind parameter $\lambda,$ and the infall mass $M_\mathrm{inf}$. 
Moreover, the range of values at which 75\% of the galaxies are found is presented.  
Results are also indicated for two different IMFs: a \citet{chabrier2003} and \citet{salpeter1955} IMF.}
\label{table}
\end{table*}

\section{Model}

In this section we present new analytical solutions for the evolution of the metallicity, gas mass, and 
total mass of galaxies in the framework of simple models of chemical evolution where an exponential infall of 
gas is assumed. 
We summarize the basic assumptions of our model and show the system of differential equations that are to be solved. 
We show the new analytical solution and explore the effect of varying the main free parameters of our model.

\subsection{Basic assumptions}

The main assumptions of the simple model \citep{tinsley1980} are as follows:

\begin{enumerate}

\item The IMF is constant in time and space, which means that every galaxy stellar generation 
hosts stars with a mass sampling a universal distribution, regardless
of the age, metallicity, and birthplace of the stellar generation. 

\item The gas is well mixed at any time of the galaxy evolution (\textup{{\it instantaneous mixing approximation}}).

\item Stars with mass $m \ge1\,\mathrm{M}_{\sun}$ die instantaneously, as soon as they form 
(IRA), while stars with mass $m<1\,\mathrm{M}_{\sun}$ have infinite lifetimes. 

\end{enumerate}

\noindent
By making these simplifying assumptions, we can derive analytical formulas for the evolution of the 
main galaxy physical properties, such as the metallicity $Z$, SFR, gas, and stellar mass;  
the two following quantities appear in the various differential equations: 

\begin{equation}
R = \int_1^\infty (m - M_R) \phi (m) dm,
\label{eq:r}
\end{equation}
\noindent
which represents the so-called returned mass fraction, where $\phi (m)$ is the IMF and $M_{R}$ is the mass of the stellar remnant, and 

\begin{equation}
y_Z = {1 \over {1 - R}} \int_1^\infty m\,p_{Z}(m)\,\phi (m) dm,
\label{eq:yield}
\end{equation}
\noindent
which represents the so-called yield per stellar generation, with
$p_{Z}(m)$ being the fraction of the newly produced and ejected metals
by a star of mass $m$. The values of $y_Z$ and $R$ for different IMFs
are taken from \citet{vincenzo2016}, who also showed that the effect
of metallicity on these quantities is minor compared to the adopted IMF and set of stellar yields. 
 The IMF is defined in the stellar mass range 0.1- 100 M$_{\odot}$.
In
particular, we assume average values over metallicity from
Table 2 of
\citet{vincenzo2016}, corresponding to the compilation of stellar yields of \citet{romano2010}. 

\par In our model we explore the effect of the two following IMFs: 
\begin{enumerate}
\item \citet[which is similar to \citealt{kroupa2001}]{chabrier2003}, for which we obtain a return mass fraction $R=0.441$, a yield of metals per stellar generation $y_Z=0.0631,$ 
and a yield of oxygen per stellar generation $y_O = 0.0407$; 
\item  \citet{salpeter1955}, for which $R=0.287$, $y_Z=0.0301,$ and  $y_O=0.018$.
\end{enumerate}

We assume for the SFR a linear \citet{schmidt1959} law of the
following form:
\begin{equation}
\psi(t)= S \times M_\mathrm{gas}(t),  
\end{equation} 
where $M_\mathrm{gas}(t)$ is the galaxy gas mass at the time $t,$ and $S$ is the so-called star formation efficiency (SFE), 
a free parameter of our model that is measured in $\mathrm{Gyr}^{-1}$.

 The infall model is taken from \citet{chiosi1980} and the rate is assumed to obey the following decaying exponential law: 
\begin{equation}
I(t) = A e^{-t/{\tau}},
\label{eq:I}
\end{equation}
where $\tau$ is the so-called infall timescale, which determines the typical timescale over which the galaxy is assumed to assemble, 
and $A$ is a constant constrained by the total infall gas mass ($M_\mathrm{inf}$) by the following equation:  

\begin{equation}
\int_0^{t_\mathrm{G}} A e^{-t/{\tau}}  dt=M_\mathrm{inf}  \quad \Rightarrow \, A=\frac{M_\mathrm{inf} }{\tau\big(1-e^{-t_\mathrm{G}/\tau}\big)}.
\label{cond}
\end{equation}
\noindent
With Eq. (\ref{cond}) we impose that the integral of the infall rate
over the entire galaxy lifetime is the total gas infall mass,
$M_\mathrm{inf}$. We assume the galactic lifetime to be
$t_\mathrm{G}=14\,\mathrm{Gyr}$.

In our model, we also take into account outflow gas episodes in galaxies. The outflow rate is assumed to be proportional to the SFR in the galaxy (see \citealt{matteucci2012,matteucci1983}):
\begin{equation}
W (t) = \lambda \psi (t),
\label{eq:w}
\end{equation}
with the wind parameter $\lambda$ being a dimensionless quantity.  We
 are aware that in our study galactic winds are treated in a simple
 way.  On the other hand, \citet{recchi2013} showed that the pure
 galactic wind mass-loss depends on the gas stratification: we can
 obtain various wind strengths and different mass
 loading factors for the same SFR.  However, Eq. (\ref{eq:w}) has been used in almost
 all the chemical evolution models in the literature up to now
  (see \citealt{vincenzo2014, recchi2008,lanfranchi2008}).
\subsection{Set of differential equations}
 
 The set of differential equations we have to solve to characterise the evolution of the galaxy total mass, gas mass, and metallicity is the 
 following:

\begin{equation}
\small{
\begin{cases}
 \dot{M}_\mathrm{tot}(t) =  A e^{-t/{\tau}} -\lambda \psi(t) & \\[7pt] 
 \dot{M}_\mathrm{gas}(t) =  -\big(1 - R \big) \psi(t) +  A e^{-t/{\tau}}- \lambda \psi(t)  & \\[7pt]
\dot{M}_Z(t) = \big(-Z(t)+y_z \big)\big(1-R\big) \psi(t) -\lambda Z(t) \psi(t) + Z_\mathrm{inf} A e^{-t/{\tau}} 
\label{system1}
\end{cases} }
,\end{equation}
where $M_\mathrm{tot}$ and $M_\mathrm{gas}$ are the total mass and the gas mass of the galaxy, respectively; 
 $Z = M_Z/M_\mathrm{gas}$ represents the gas metallicity of the system, and 
$Z_\mathrm{inf}$ the metallicity of the infalling gas. 
The total stellar mass of the system can be retrieved by means of the following formula: $ M_{\star}(t)=M_\mathrm{tot}(t)-M_\mathrm{gas}(t)$.

By recalling that $M_Z = Z\times M_\mathrm{gas}$, if we differentiate the latter with respect to time and then 
combine the second and the third equations in the system of Eq. (\ref{system1}), 
the temporal evolution of the gas metallicity, $Z$, obeys the following differential equation: 

\begin{equation}
\dot{Z}(t) =y_z\big(1-R\big)S + \frac{A\big(Z_\mathrm{inf}-Z(t)\big)e^{-t/\tau}}{M_\mathrm{gas}(t)}. 
\label{ets}
\end{equation}
By assuming $Z_\mathrm{inf} = 0$, the second term in the right-hand side of Eq. (\ref{ets}) 
is responsible for a dilution of the metal content within the galaxy ISM as the infall rate of gas proceeds. 

\subsection{Analytical solutions}
\label{sec:analytical_solutions}
In this section, we present the analytical solutions for the system of Eq. (\ref{system1}) 
and show the effects on $M_\mathrm{gas}(t)$,  $M_\mathrm{tot}(t)$, and $Z(t)$ 
caused by the variation of the adopted IMF and wind parameter $\lambda$. 

\par We assume for Eq. (\ref{system1}) the following initial conditions: 
\begin{enumerate}
 \item At $t=0$, we assume $M_\mathrm{tot}(0)=M_\mathrm{gas}(0) \ll M_\mathrm{inf}$. 
 To integrate Eq. (\ref{system1}), an initial gas mass different from zero is required; 
 we consider it as negligible with respect to the infall mass.  
\item  The metallicity of the gas infall is constant: $Z_\mathrm{inf} = 0$.\item  The metallicity of the galaxy is primordial at the formation of the galaxy: $Z(0)=Z_\mathrm{inf}=0$.
\end{enumerate}

The solution for the gas mass is 
\begin{equation}
 M_\mathrm{gas}(t)= e^{-\alpha t} \left( \frac{ A  \left[ 
           e^{-t/\tau + \alpha  t} - 1 \right]  \tau }{ \alpha \tau - 1}  +     M_\mathrm{gas}(0)           \right),\end{equation}
\noindent
while the total galaxy mass ($ M_\mathrm{gas}(t)+M_\mathrm{\star}(t)$) evolves according to the following formula: 
\begin{equation}
\begin{split}
  M_\mathrm{tot}(t) & = \; M_\mathrm{gas}(0)\, \frac{S}{\alpha} \, \left(\lambda e^{-\alpha t} +1-R       \right) + \\[4pt]
 + \, & \frac{S}{\alpha}\,\big(1 - R\big) A  \tau - A\tau  e^{-t/\tau }
+ \frac{S}{\alpha} \frac{ \lambda A   \tau \big( \tau \alpha e^{-t/\tau } -e^{-\alpha t}  \big) }{\big(  \alpha  \tau-1\big)} \nonumber.
\end{split}
\end{equation}

\noindent
Finally, the solution for the galaxy gas-phase metallicity is 
\begin{equation}
\begin{split}
 Z(t)  =  \; & \frac{y_z  S\big( 1-R \big)}{ \alpha \tau-1}  \times \\[4pt] 
\times \, & \frac{M_\mathrm{gas}(0) t \big( \alpha \tau - 1 \big)^2+ A \tau\big[t - \tau  (1  + \alpha  t)    +  \tau   e^{\alpha t -t/\tau}  \big]   }
{  A \tau  \big(e^{\alpha t -t/\tau} -1\big) + M_\mathrm{gas}(0)\big( \alpha \tau-1\big) } \nonumber . 
\label{Znew}
\end{split}
\end{equation}

\noindent
In these equations, we have defined the parameter $\alpha$ as follows:
\begin{figure}
\includegraphics[scale=0.7]{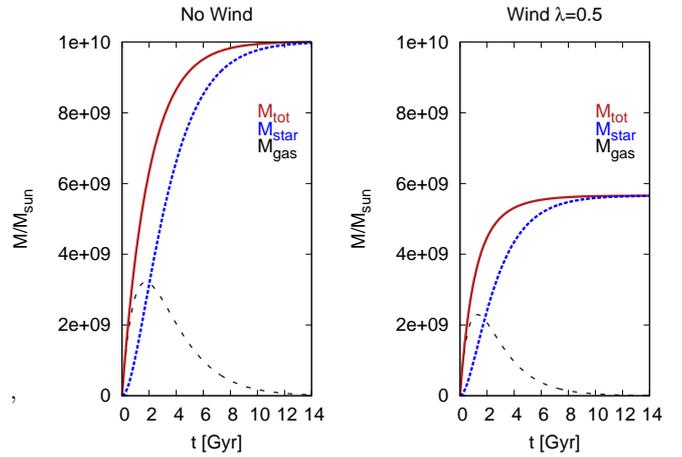}
 \caption{Effects of galactic winds. We show in both panels the predicted evolution in time 
 of the galaxy gas mass $M_\mathrm{gas}$ (gray dashed lines), stellar mass $M_{\star}$ (blue dotted lines), 
 and total mass $M_\mathrm{tot} = M_\mathrm{gas} + M_{\star}$ (red solid lines). 
 The model assumes an exponential infall law with timescale $\tau=2$ Gyr and infall mass   
 $M_\mathrm{inf}=10^{10}\,\mathrm{M}_{\odot}$, and an SFE $S=1\,\mathrm{Gyr}^{-1}$. 
 {\it Left panel}: model without galactic winds  ($\lambda=0$); {\it right panel}: model with wind parameter $\lambda=0.5$. }
 \label{newsol2panel}
\end{figure} 
\begin{figure}
          \centering \includegraphics[scale=0.7]{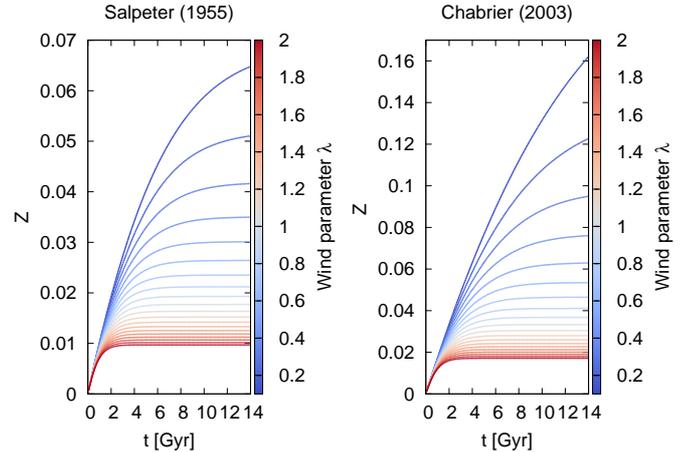} 
\caption{Effect of varying the wind parameter $\lambda$ on the time evolution of the metallicity $Z$. 
The model assumes an exponential infall law with timescale $\tau=2$ Gyr and infall mass   
 $M_\mathrm{inf}=10^{10}\,\mathrm{M}_{\odot}$, and an SFE $S=1\,\mathrm{Gyr}^{-1}$. In the left panel we show the 
 results of models with a \citet{salpeter1955} IMF, while in the right panel a \citet{chabrier2003} IMF is assumed. }
 \label{Znewfig}
\end{figure} 

\begin{equation}
\alpha= \big(1 + \lambda - R\big)S.
\end{equation}
We note that when we consider the case without infall mass ($M_\mathrm{inf}=0$, and hence $A=0$) and without winds ($\lambda=0$), we recover the closed-box solution.
 We  have with these assumptions that 

\begin{equation}
 M_\mathrm{tot,cb}(t)=    M_\mathrm{gas}(0),
\end{equation}
\begin{equation}
 M_\mathrm{gas,cb}(t)=    M_\mathrm{gas}(0) e^{-(1-R)St},
\end{equation}

and

\begin{equation}
 Z_{cb}(t)=y_z  S (1-R) t.
\end{equation}

$ M_{tot,cb}(t)$, $M_{gas,cb}(t),$ and  $Z_{cb}(t)$ are exactly the solutions of the closed-box model reported in \citet{spitoni2015} for the total mass, gas mass, and metallicity of the gas, respectively.

In Fig. \ref{newsol2panel} we show how $M_\mathrm{gas}$, $M_{\star}$ and $M_\mathrm{tot}$ 
evolve when we assume an infall timescale $\tau=2\,\mathrm{Gyr}$, infall mass 
$M_\mathrm{inf}=10^{10}\,\mathrm{M}_{\sun}$, 
and SFE $S=1\,\mathrm{Gyr}^{-1}$. 
A model with wind parameter $\lambda=0.5$ (right panel) is compared with a model without outflow (left panel). 
As the time tends to $t_\mathrm{G}$, both models predict the total galaxy gas mass to be zero and hence the stellar mass to 
approach the galaxy total mass. 
In the model without galactic winds, $M_\mathrm{\star}(t_\mathrm{G}) \approx M_\mathrm{tot}(t_\mathrm{G}) = M_\mathrm{inf}$, while 
in the model with outflow we find that 
$M_\mathrm{\star}(t_\mathrm{G}) \approx M_\mathrm{tot}(t_\mathrm{G}) < M_\mathrm{inf}$, since 
the galaxy loses a substantial fraction of its total infall mass  
in the intergalactic medium (IGM) because of the galactic winds. 
The shape of the temporal evolution of the galaxy SFR can be retrieved from the behavior of 
$M_\mathrm{gas}(t)$, since we assume them to be proportional between each other.

\begin{figure}
          \centering \includegraphics[scale=0.7]{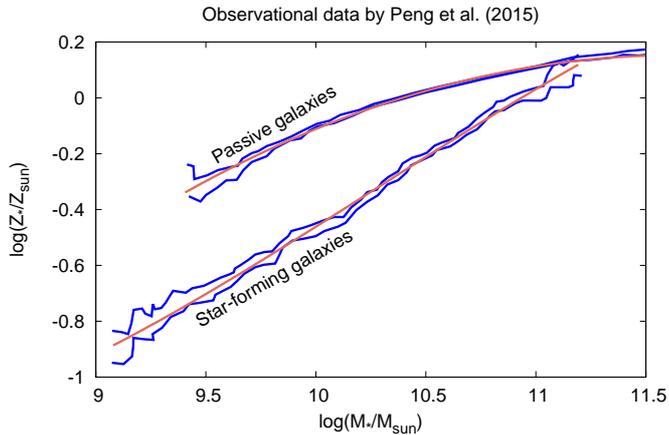} 
\caption{Observed  uncertainties of the MZ relations for star-forming and passive galaxies by \citet{peng2015} are reported with blue lines. We show with orange lines the  third-order polynomial fits  for the passive and star-forming galaxies  adopted in this article. }
 \label{obs}
\end{figure}

\begin{figure*}
          \centering
\includegraphics[scale=0.67]{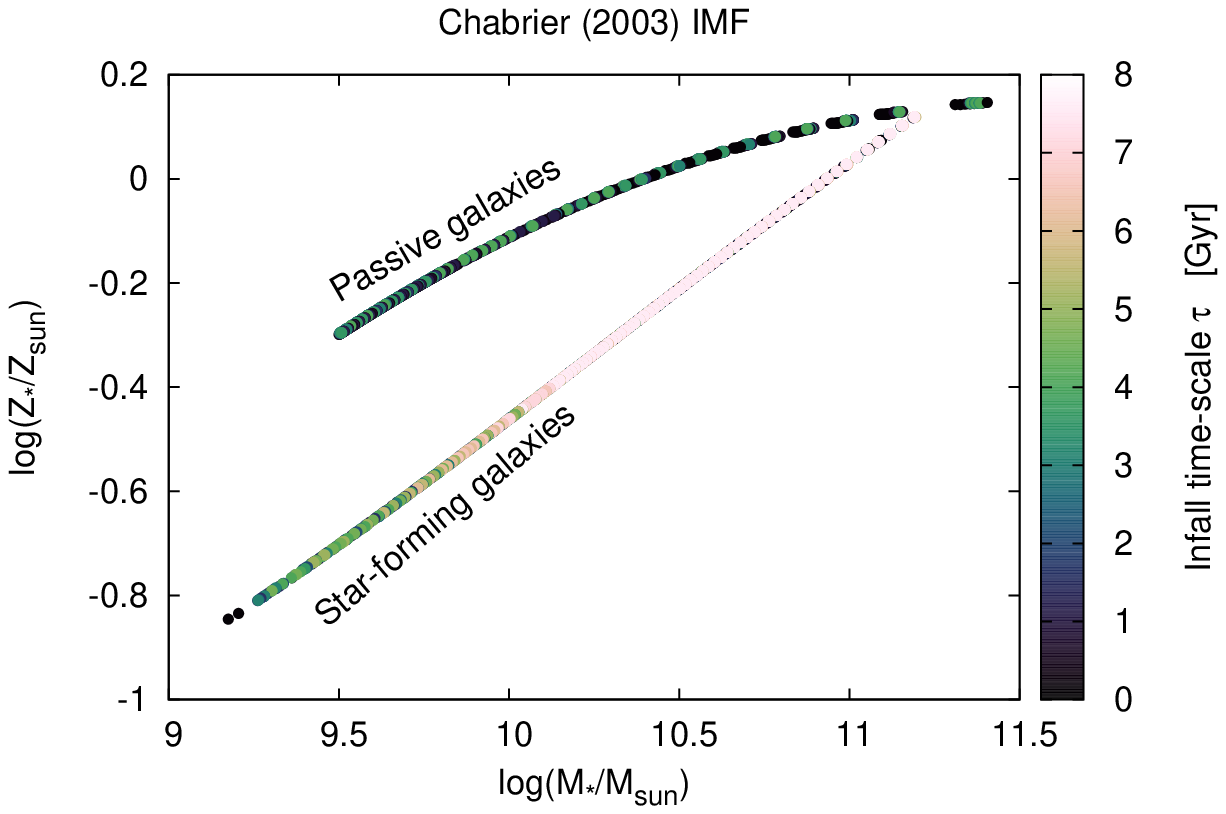}
\includegraphics[scale=0.67]{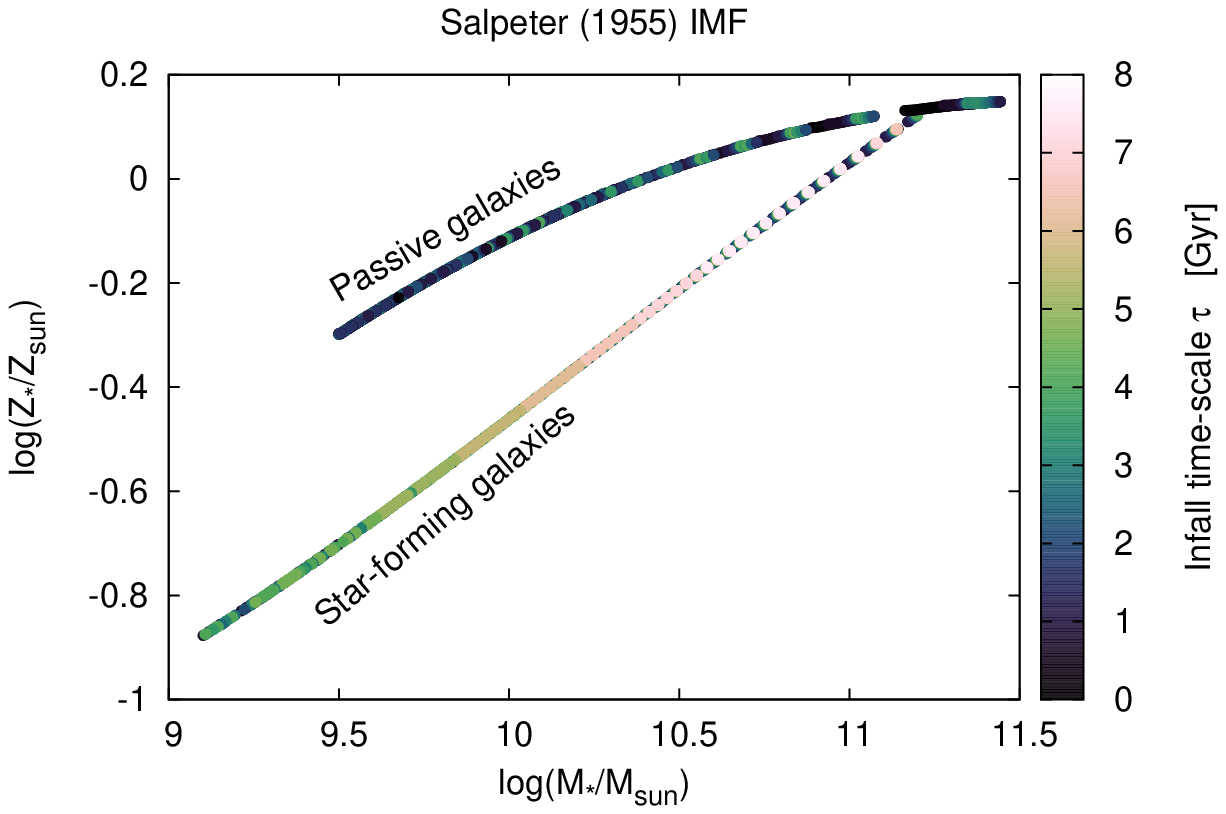}
\includegraphics[scale=0.45]{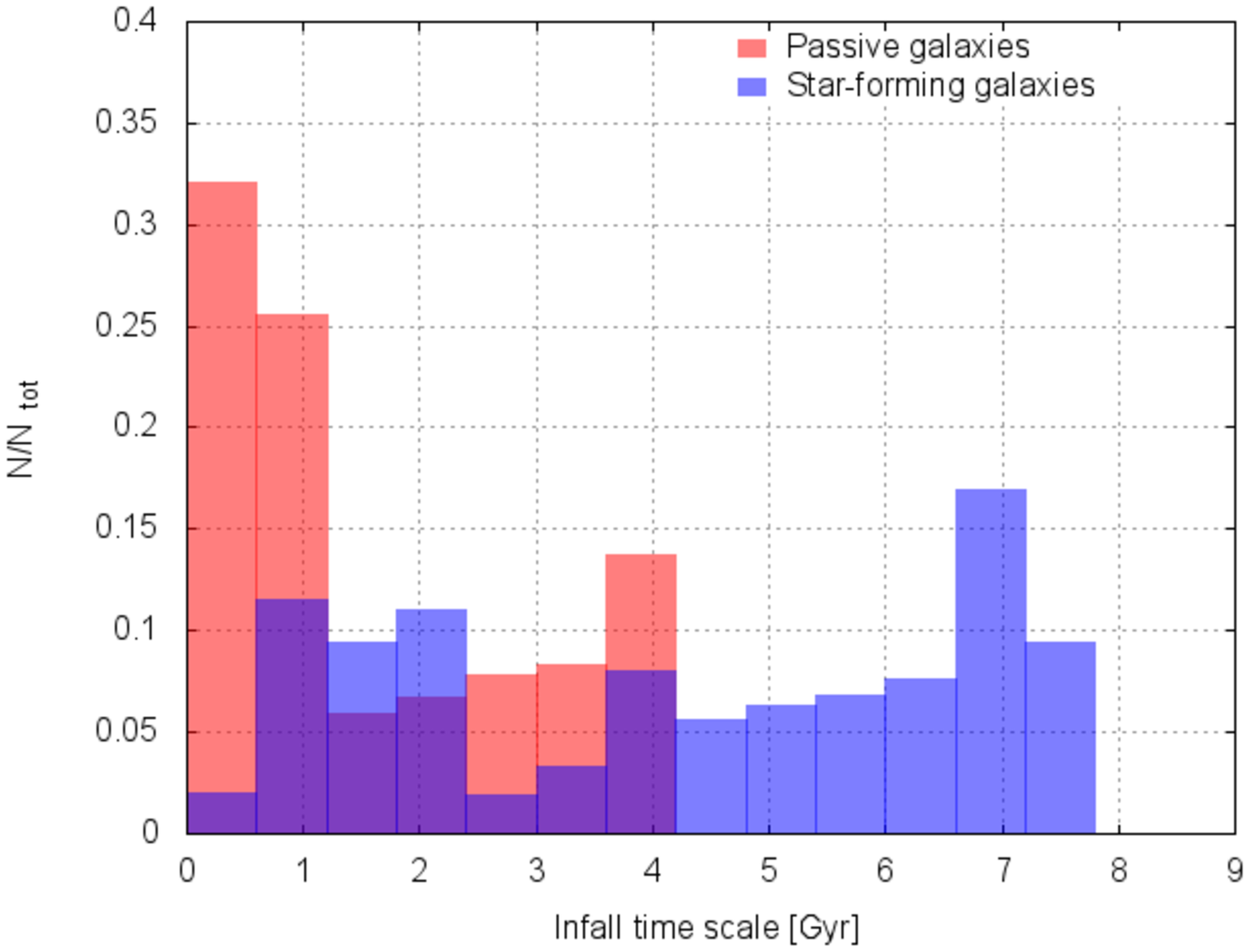}
\includegraphics[scale=0.45]{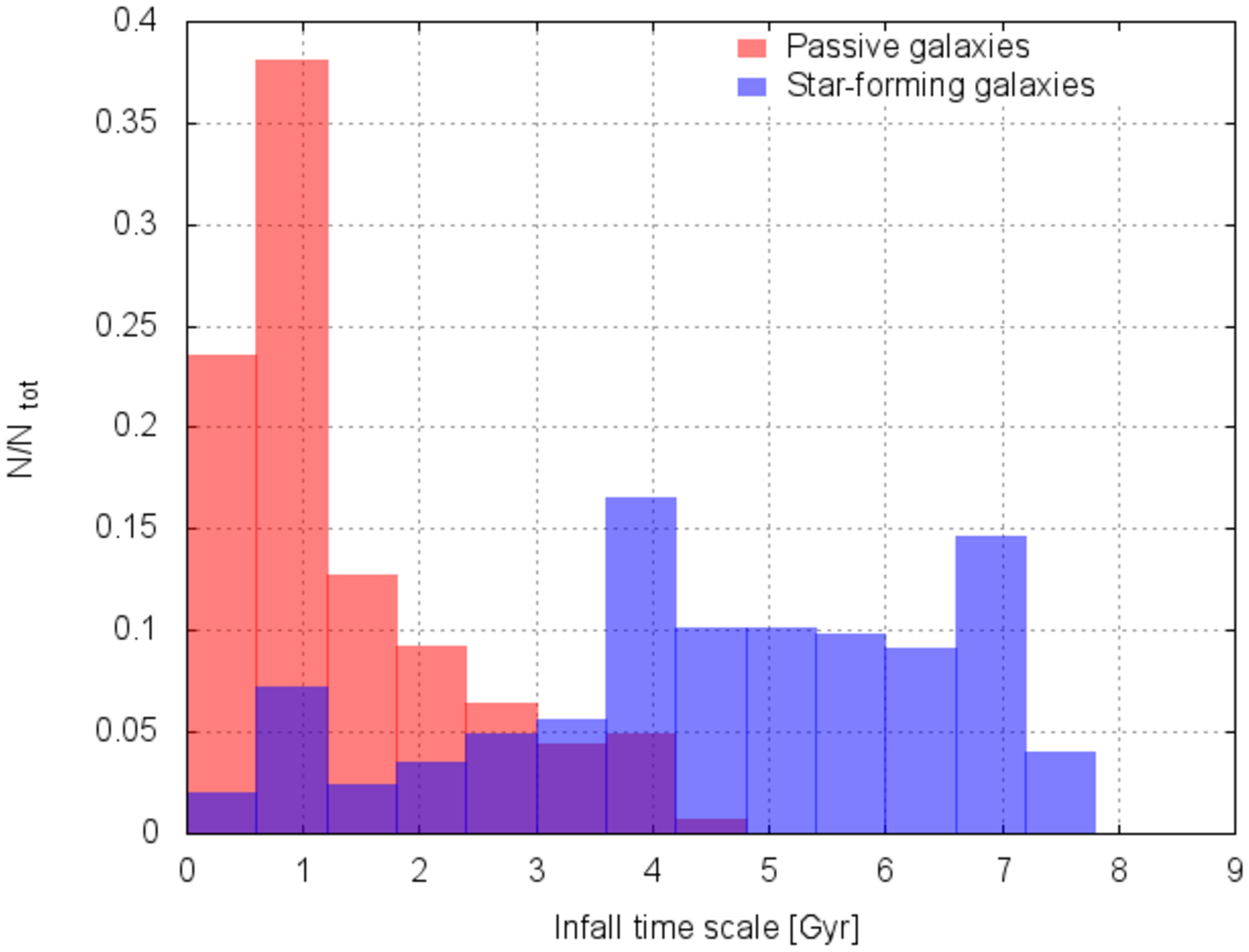}
\caption{ {\it Upper panels}: The MZ relation by \citet{peng2015} (fitted by  third-order polynomial functions shown in Fig. \ref{obs}) for passive and star-forming galaxies.  We  reproduce it  using the new analytical solution presented in this paper and vary different parameter models. In this figure  we focus on the timescale parameter $\tau$, and the color code indicates different  $\tau$ values  of   the computed galaxies that reside along  the  MZ relations. In the left panel we adopt a \citet{chabrier2003} IMF, whereas in the right panel the \citet{salpeter1955} IMF
is assumed. {\it Lower panels}: The distribution of the predicted passive galaxies (red histogram) and star-forming galaxies (blue histogram)  that reproduced the MZ relation in terms of the timescale parameter $\tau$ with a \citet{chabrier2003} IMF (left panel) and a \citet{salpeter1955} IMF (right panel). }    \label{TD}
\end{figure*}

In Fig. \ref{Znewfig} we show how the gas metallicity $Z$ is
predicted to evolve with time when we assume different wind parameters,
$\lambda$. The right panel shows the results of models with a
\citet{chabrier2003} IMF, while the left panel shows models with a \citet{salpeter1955} IMF. All the models assume 
$M_\mathrm{inf}=10^{10}\,\mathrm{M}_{\odot}$, $\tau=2\,\mathrm{Gyr,}$
and $S=1\,\mathrm{Gyr}^{-1}$.  The figure clearly shows
that as the wind parameter $\lambda$ increases, the galaxy metallicity
is predicted to saturate at progressively earlier epochs toward always
lower metallicities.  Interestingly, the assumption of
a \citet{chabrier2003} IMF causes an enhanced chemical enrichment of
the galaxy ISM. At any galactic time, the gas metallicity
with a \citet{chabrier2003} IMF is about twice higher
than the metallicity with a \citet{salpeter1955} IMF, which is due to the
large portion of massive stars in the \citet{chabrier2003} IMF.  As
expected, the choice of the IMF has a strong effect on the metallicity
evolution of a galactic system.

\begin{figure*}
          \centering 
  \includegraphics[scale=0.67]{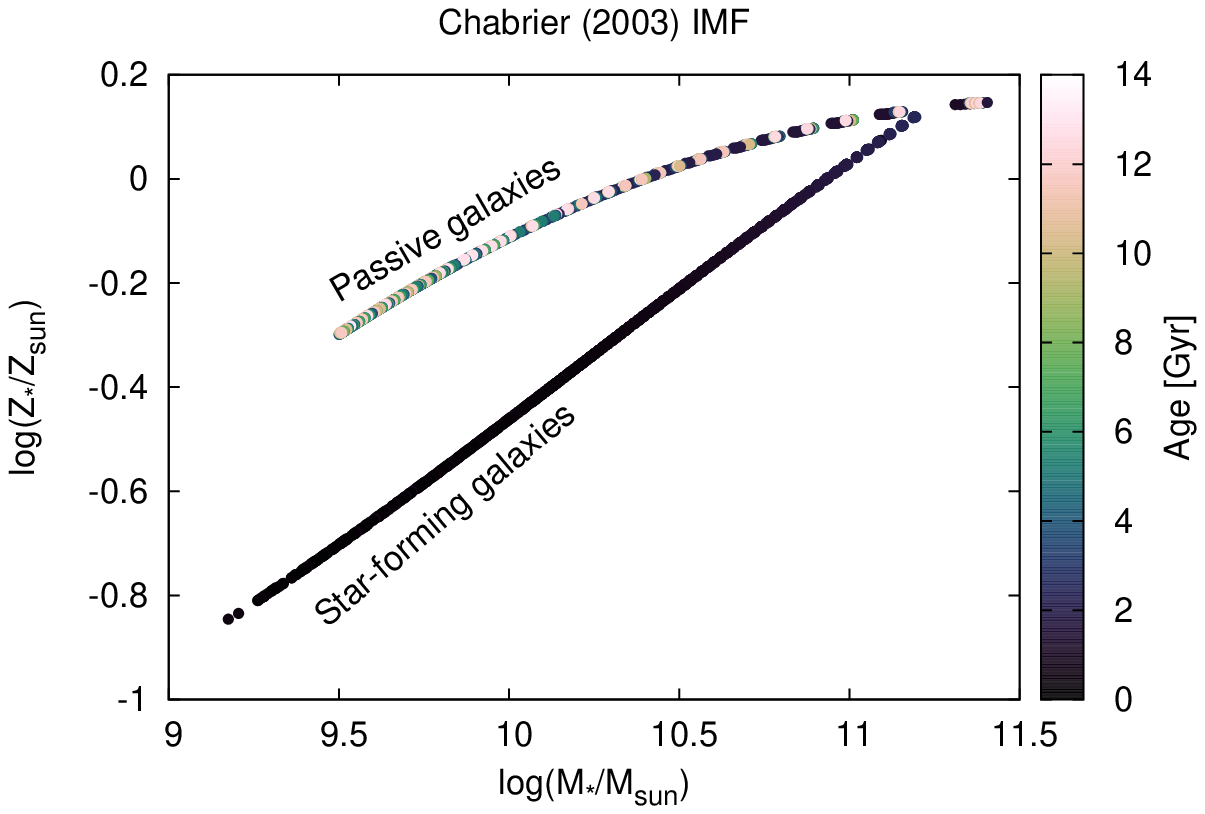} 
 \includegraphics[scale=0.67]{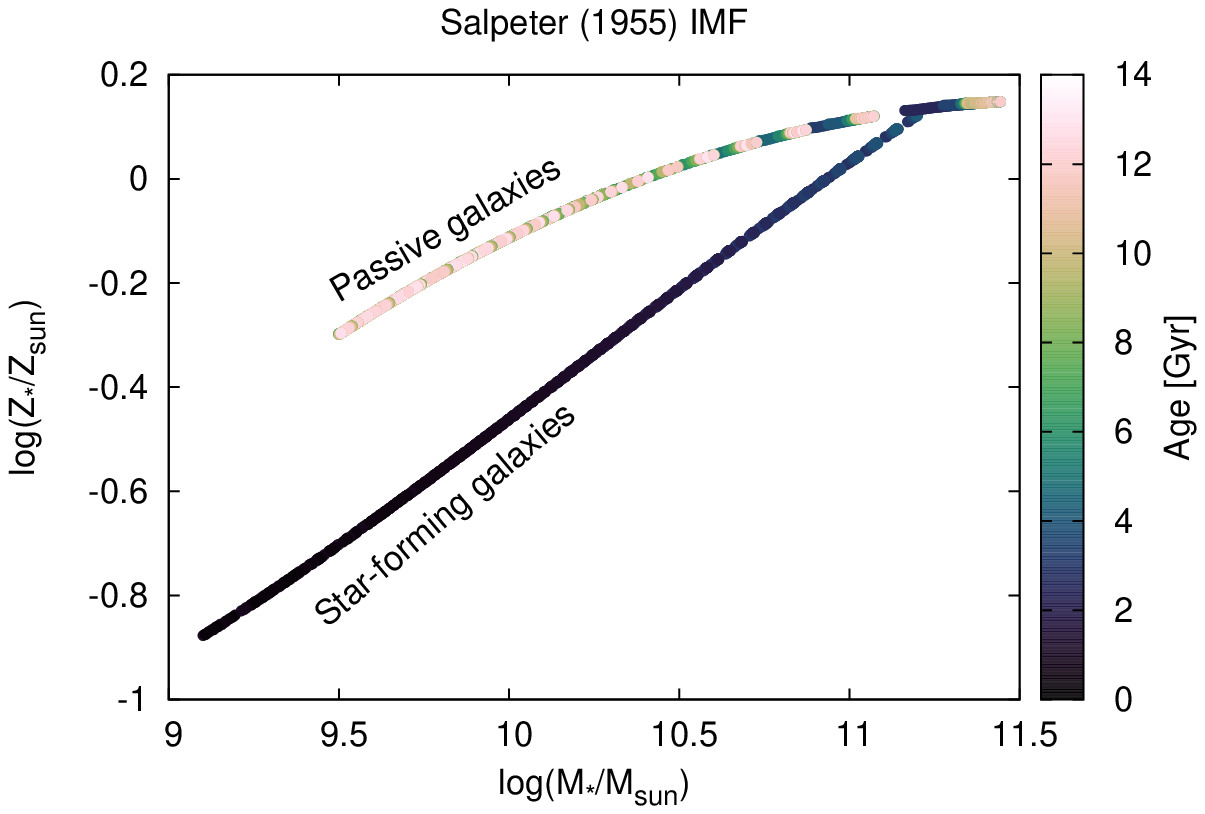} 
\includegraphics[scale=0.45]{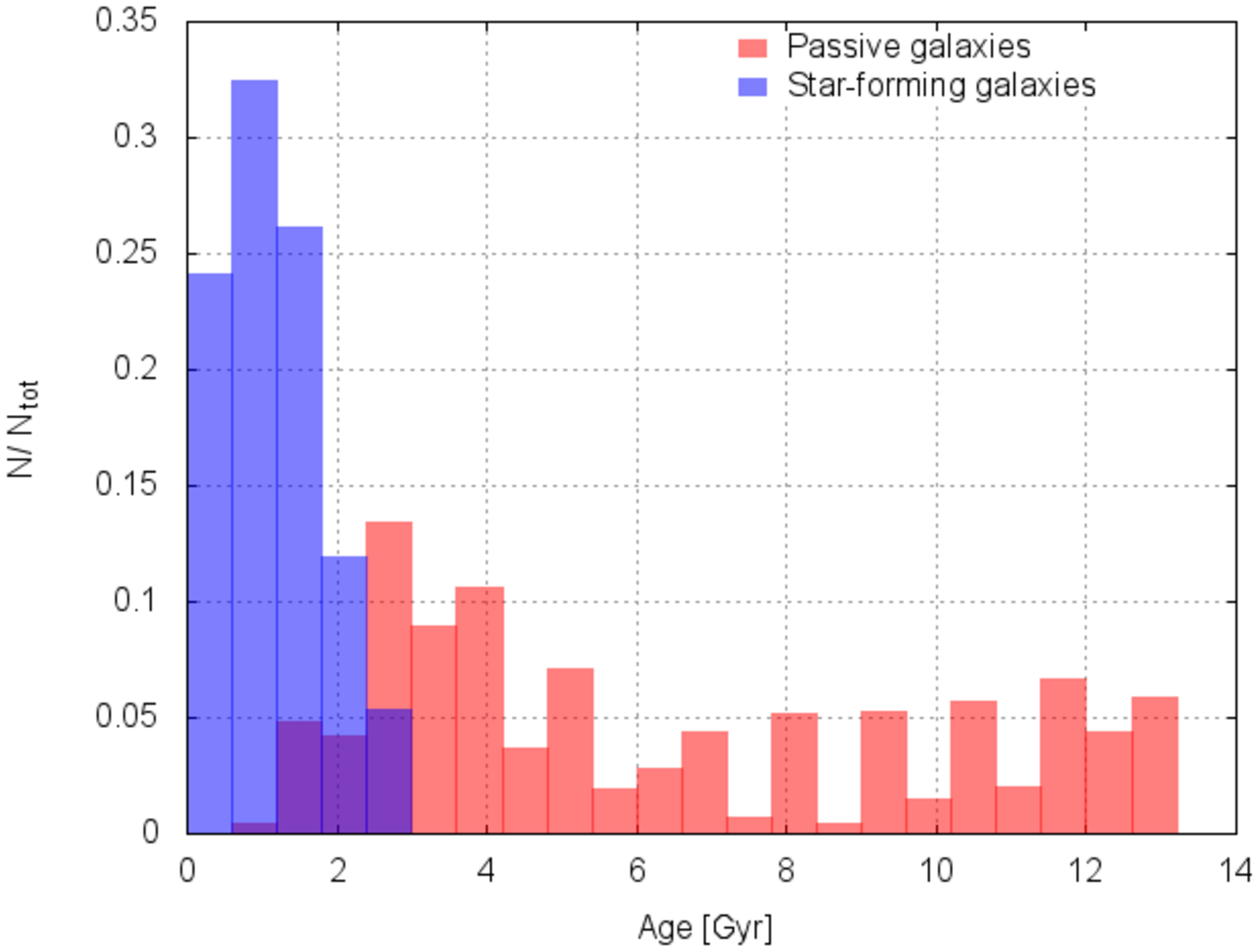} 
\includegraphics[scale=0.45]{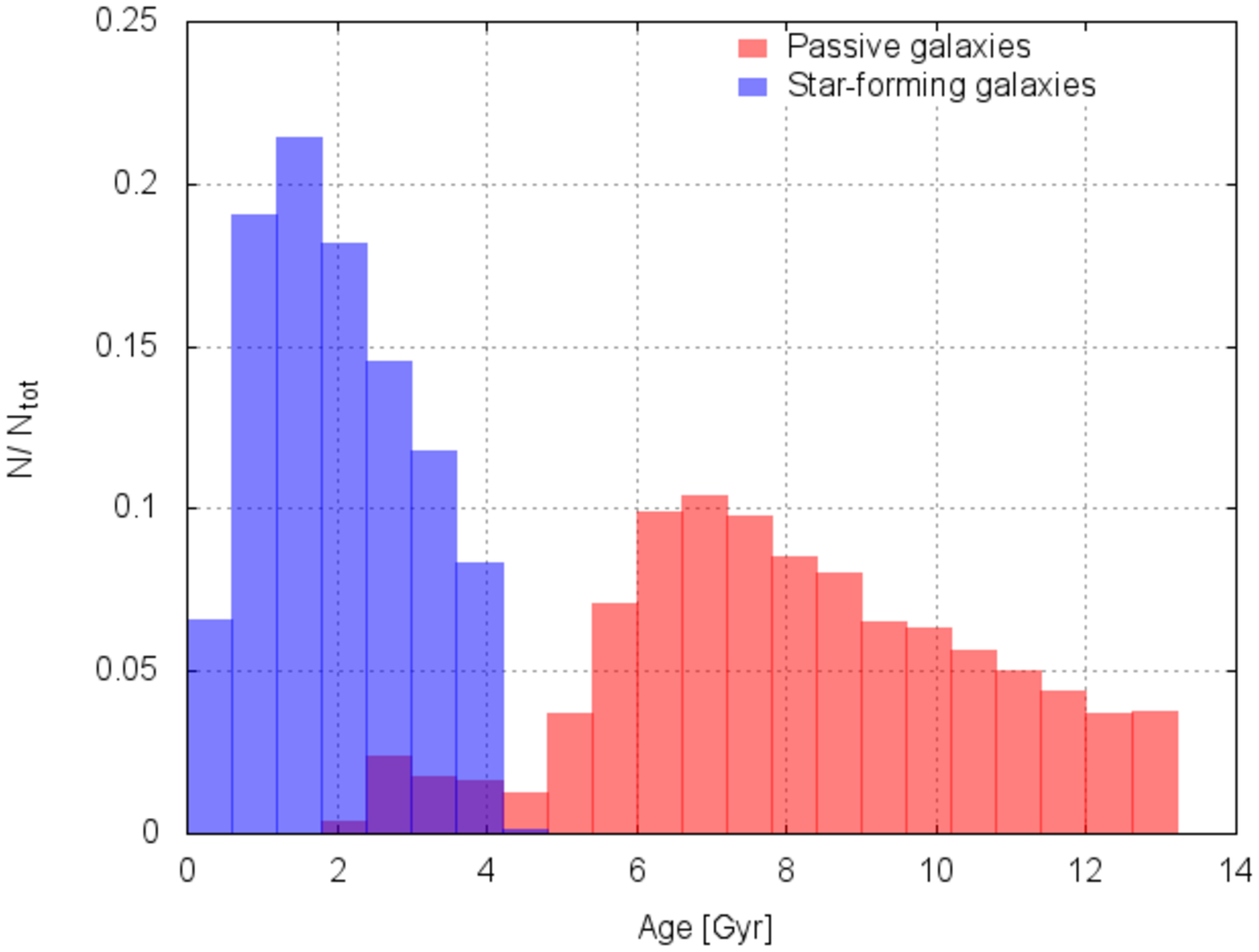}
        \caption{{\it Upper panels}: As in Fig. \ref{TD}, but here, the
color code indicates different ages of the galaxies that reproduce
the MZ relations.  {\it Lower panels}: The distribution of the
predicted passive galaxies (red histogram) and star-forming galaxies (blue
histogram) that reproduced the MZ relation in terms of the age of galaxies.  In the left panel we adopt a Chabrier (2003) IMF, whereas in the right panel we show the results assuming a Salpeter (1955) IMF.}
\label{age}
\end{figure*}

\section{MZ relation of the SDSS sample of star-forming and passive galaxies}

By analyzing a sample of local galaxies from SDSS data, 
 \citet{peng2015} were able to separate the local population of actively star-forming
and gas-rich galaxies from the passive and gas-poor galaxies.
 In Fig. \ref{obs} we report the data by \citet{peng2015}, showing that  these two populations of galaxies present distinct
relations in the MZ plane, with the passive galaxies having on
average higher stellar metallicities than the star-forming
galaxies.

For a given stellar mass, there is a gap in metallicity between
the two galaxy populations, which is observed to diminish as the
stellar mass increases. At
$M_{\star}\approx10^{11}\,\mathrm{M}_{\sun}$ and beyond, star-forming and
passive galaxies share almost the same average stellar metallicities.
 In Fig. \ref{obs} we also show the fits of the passive and star-forming sequences  by means of third-order polynomial  
functions that we use in this paper.

To explain the observed MZ relation of the star-forming and passive galaxy populations in the SDSS data, 
\citet{peng2015} suggested that  galaxies 
ceased to accrete gas from the outside and kept forming stars only by
exhausting the remaining available cold gas reservoir within their
potential well.  In this way, as soon as the galaxy is
strangulated and stops accreting gas, the concentration of metals
in the galaxy can steeply increase, and similarly the metallicity of
all the subsequent stellar generations rises.  The average time needed for
star-forming galaxies to reach the high-metallicity stripe of passive
galaxies in the MZ relation is predicted by \citet{peng2015} to be on
the order of $\sim4\,\mathrm{Gyr}$.

We do not invoke any a priori strangulation scenario
and aim at reproducing the two observed sequences of star-forming and
passive galaxies by using our new analytical model with an
exponential infall law and by varying the important parameters of
galaxy evolution.  In particular, we characterize the local SDSS galaxies
in terms of their age, infall timescale of formation, infall mass, and
wind parameter.

\section{Methods}

\begin{figure*}
          \centering 
 \includegraphics[scale=0.67]{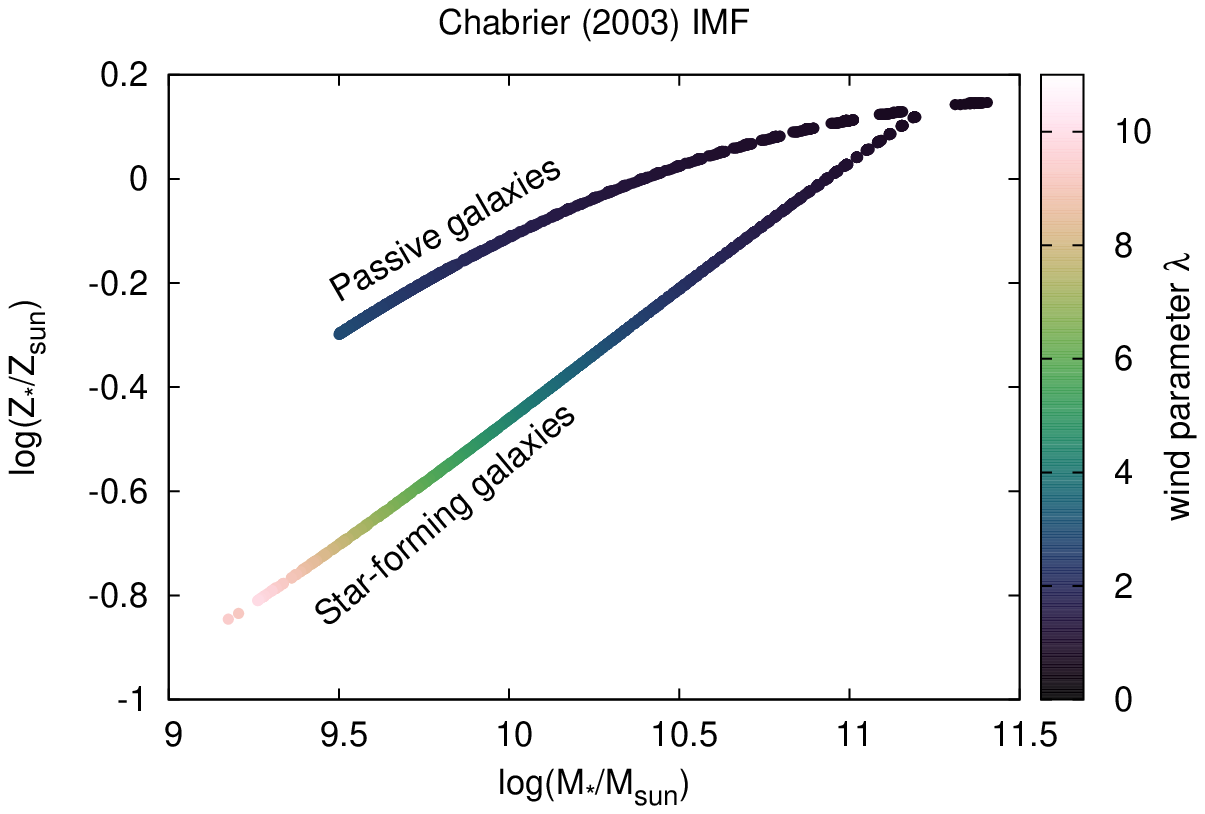}
 \includegraphics[scale=0.67]{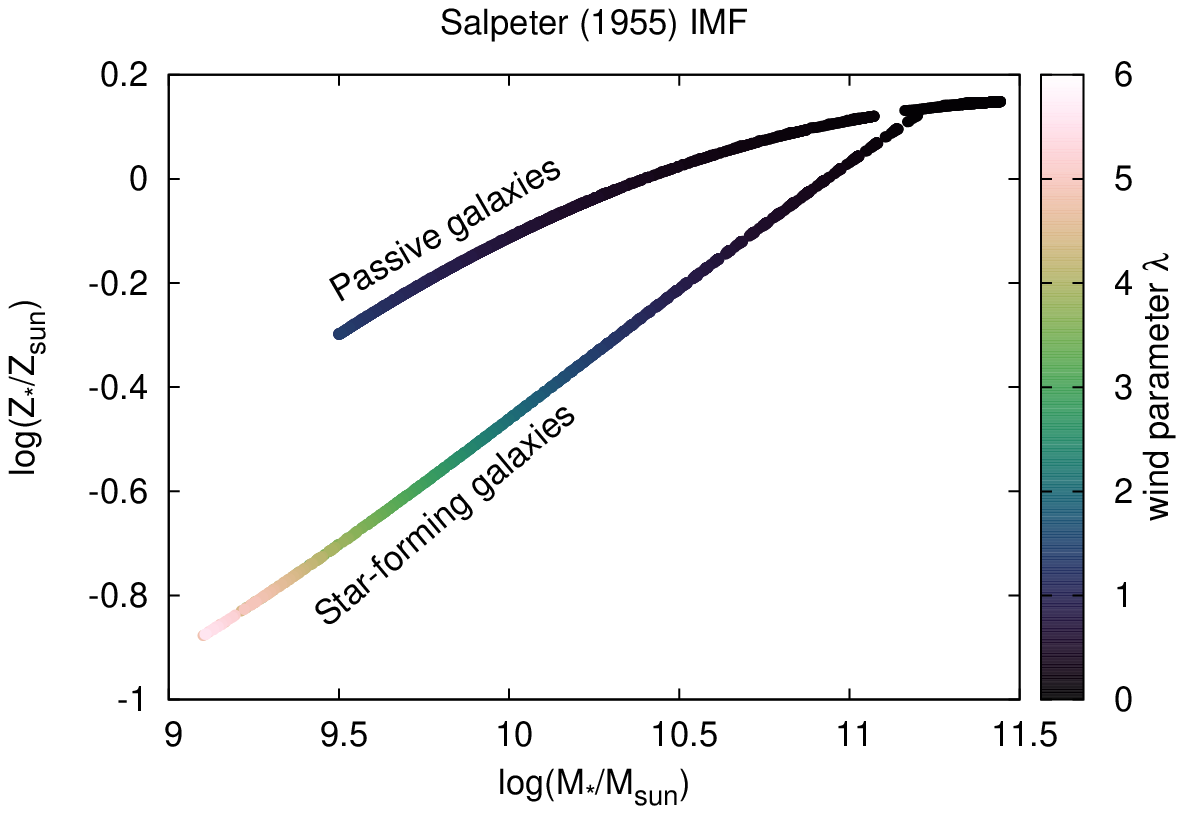}
\includegraphics[scale=0.45]{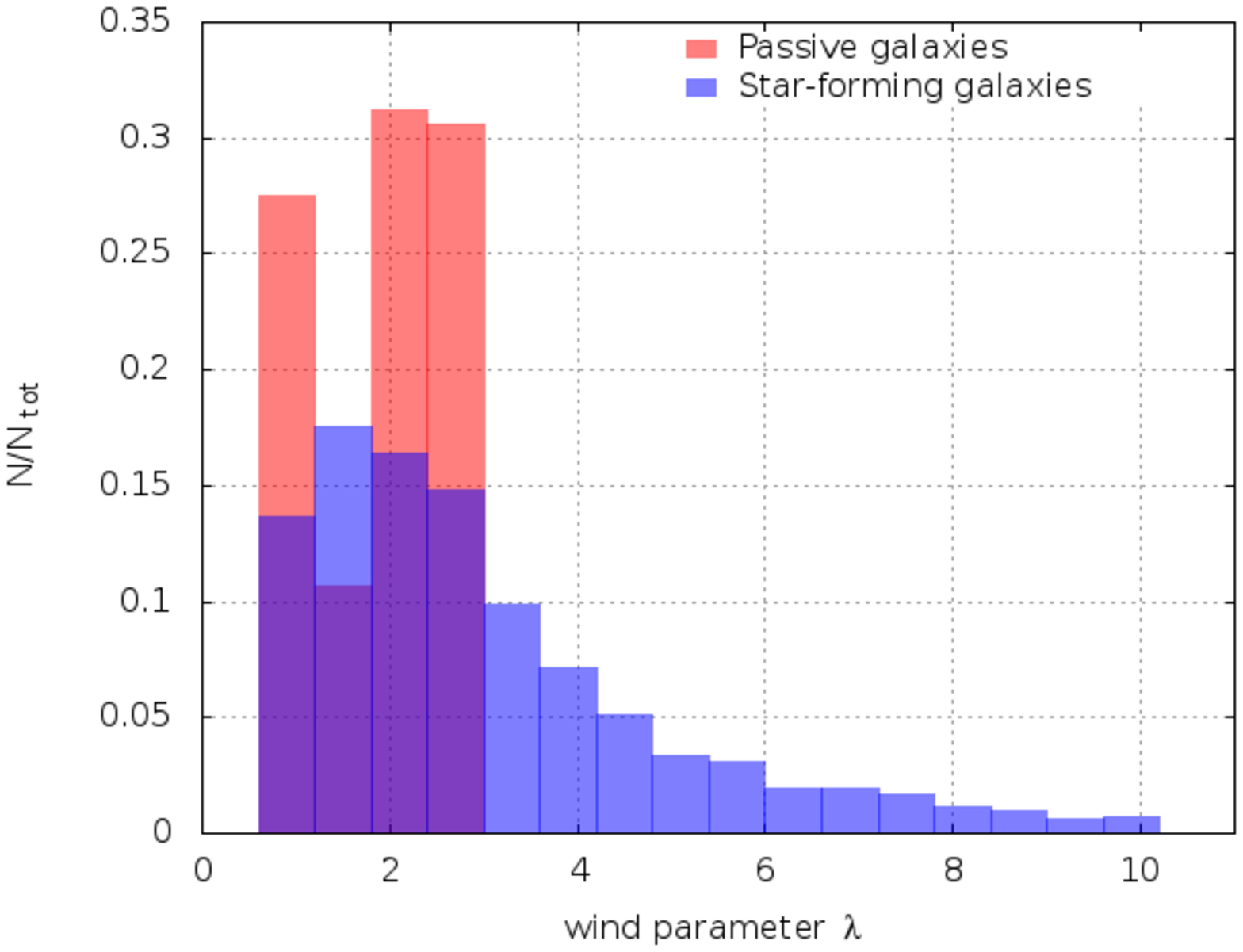} 
\includegraphics[scale=0.45]{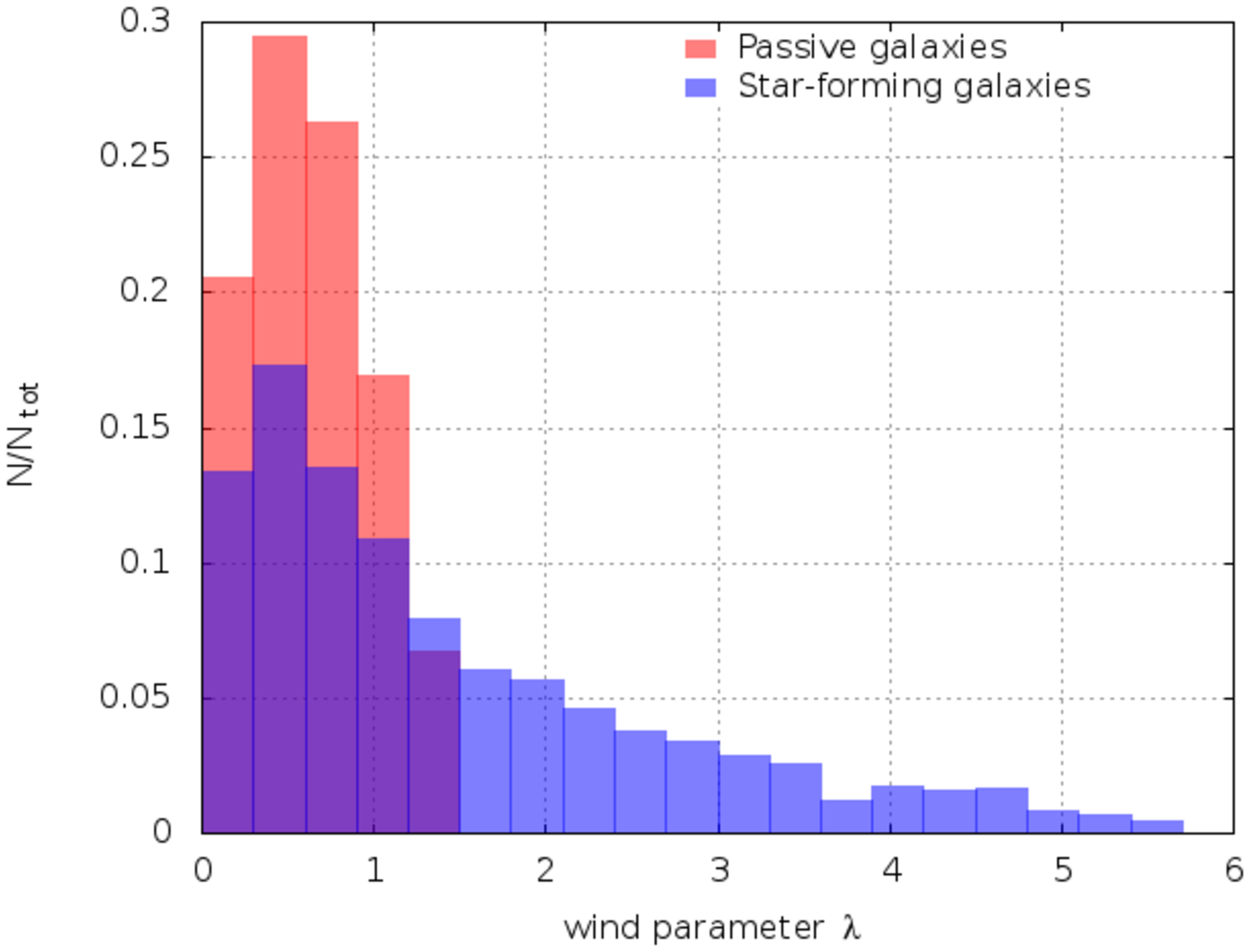} 
\caption{{\it Upper panels}: As in Fig. \ref{TD}, but here, the color
code indicates different wind parameters $\lambda$ of the galaxies
that reproduce the MZ relations. {\it Lower panels}: The distribution
of the predicted passive galaxies (red histogram) and star-forming galaxies
(blue histogram) that reproduced the MZ relation in terms of the wind
parameter $\lambda$ of galaxies.  In the left panel we adopt a Chabrier (2003) IMF, whereas in the right panel a Salpeter (1955) IMF is adopted.} \label{L}
\end{figure*}

 For the MZ relation,  \citet{peng2015} adopted the average stellar metallicity, $Z_{\star}(t)$;  
 we recover this quantity from our models by averaging the gas metallicity of each galaxy in  the following way:
\begin{equation}
\left< Z_{\star}(t) \right> =\frac{\int_0^t dt' \, Z(t') \, \psi(t')}{\int_0^t dt'\,\psi(t')}, 
\label{mass-weighted_Z}
\end{equation}
where the metallicity of the various galaxy stellar populations is weighted with the total number of stars formed with that metallicity 
(the latter quantity is directly proportional to the SFR). 
This expression represents the so-called mass-weighted average metallicity of all the stellar populations
ever born in the galaxy (see \citealt{pagel1997}). 


\subsection{Passive galaxies}

To define a galaxy as passive, we adopt the following criterium introduced by
\citet{fossati2015}, which takes into account the galaxy specific star formation rate (sSFR), that is, the SFR per unit 
galaxy stellar mass:

\begin{equation}
\mathrm{sSFR}_\mathrm{pass} < \frac{b}{t_z}, 
\label{PASSILIM}
\end{equation}
where $b$  is the birthrate parameter $b = SFR/\left<SFR\right>$ as defined
by \citet{sandage1986}, and $t_z$ is the age of the Universe at
redshift $z$. As in \citet{fossati2015}, we assume the value
proposed by \citet{franx2008}: $b=0.3$. 

Since the SDSS sample of galaxies considered by \citet{peng2015} has an average redshift $\left<z\right>=0.05$, 
the passive galaxy population is characterized by $\mathrm{sSFR}_\mathrm{pass} < 2.29 \times 10^{-11} \mathrm{yr}^{-1}$. 

\subsection{Star-forming galaxies}
For star-forming galaxies, we adopt the following scaling relation between $\mu_{\star}$ and $M_{\star}$ found by 
\citet{boselli2014} for a sample of the Herschel Reference Survey 
SDSS galaxies: 
\begin{equation}
\log(\mu_{\star})=-0.74\log( M_{\star}/M_{\sun})+7.03,
\label{mustar}
\end{equation} 
\noindent
 where $\mu_{\star}$  is the ratio between $M_{gas}$ and $M_{\star}$: $\mu_{\star}=M_{gas}/M_{\star}$.  
Molecular gas masses are estimated from the $H$-band
luminosity-dependent conversion factor of \citet{boselli2002}, while 
the galaxy stellar masses are derived from the galaxy $i$-band luminosities 
by assuming a \citet{chabrier2003} IMF and using the $g-i$ color-dependent 
stellar mass-to-light ratio relation from \citet{zibetti2009}. 

A further scaling relation found by \citet{boselli2014} is the following relation between the typical galaxy gas depletion timescale, 
$\tau_\mathrm{gas}$,  and the galaxy stellar mass:
\begin{equation}
\log(\tau_\mathrm{gas})=-0.73\log( M_{\star}/ M_{\sun})+16.75,
\label{taugas}
\end{equation}
where $\tau_\mathrm{gas}=M_\mathrm{gas}/\mathrm{SFR}$ is defined as the inverse of our SFE, namely $\tau_\mathrm{gas} = 1/S$. 
According to Eq.(\ref{taugas}), galaxies with higher stellar mass would consume their available gas mass on progressively shorter typical timescales 
if only star formation activity were taking place in the galaxy; this means that larger galaxies are expected to experience, on average, higher SFEs (see \citealt{matteucci2012}). 
We adopt Eq. (\ref{taugas}) to constrain the galaxy SFE (which is kept fixed during the galaxy evolution), 
given an initial value for the galaxy infall mass, $M_\mathrm{inf}$.

\begin{figure*}
          \centering 
\includegraphics[scale=0.67]{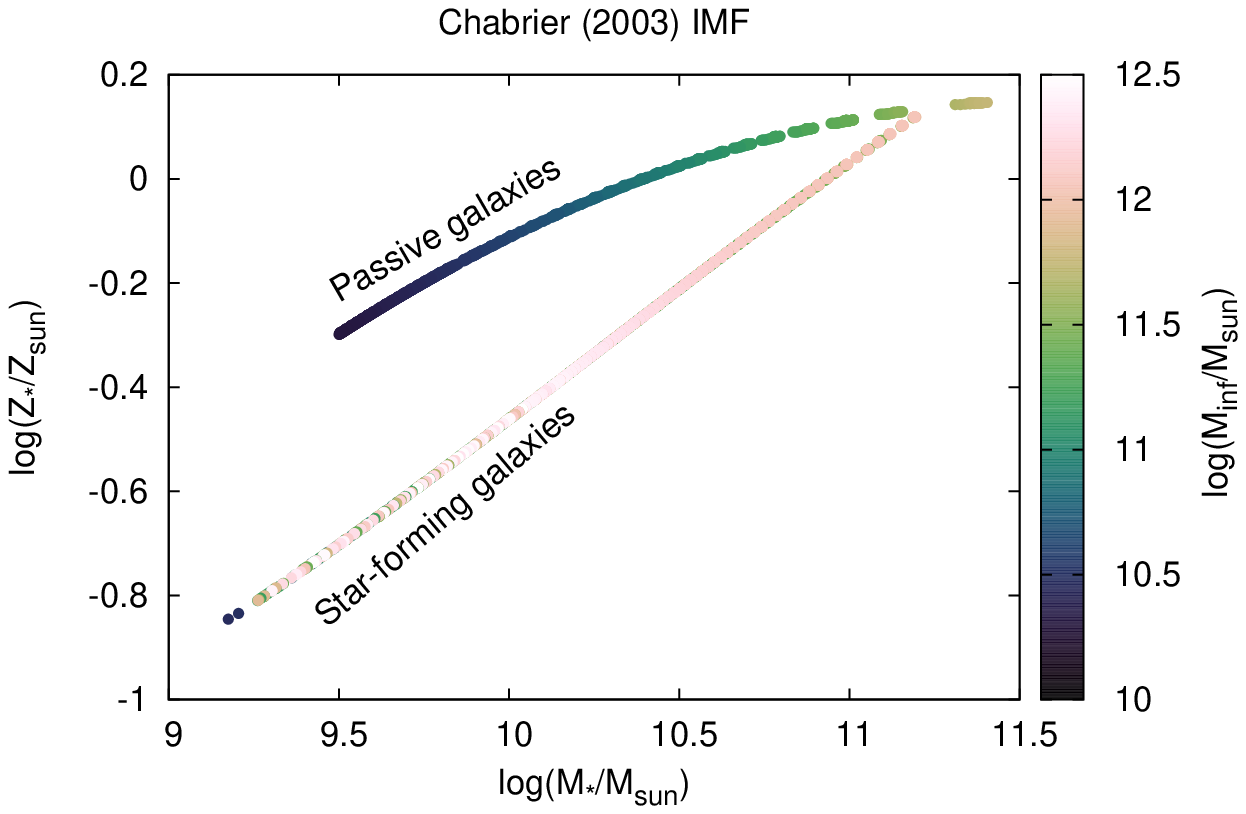}
\includegraphics[scale=0.67]{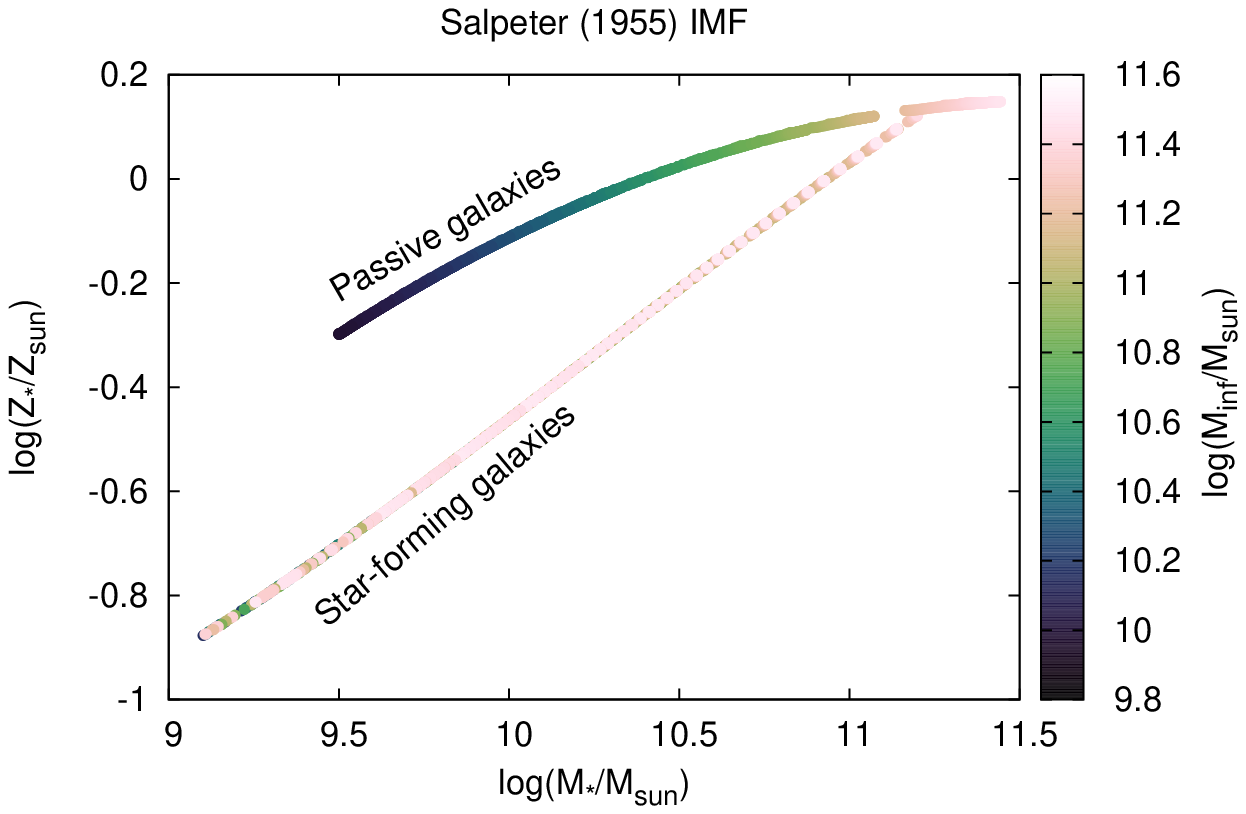}
\includegraphics[scale=0.45]{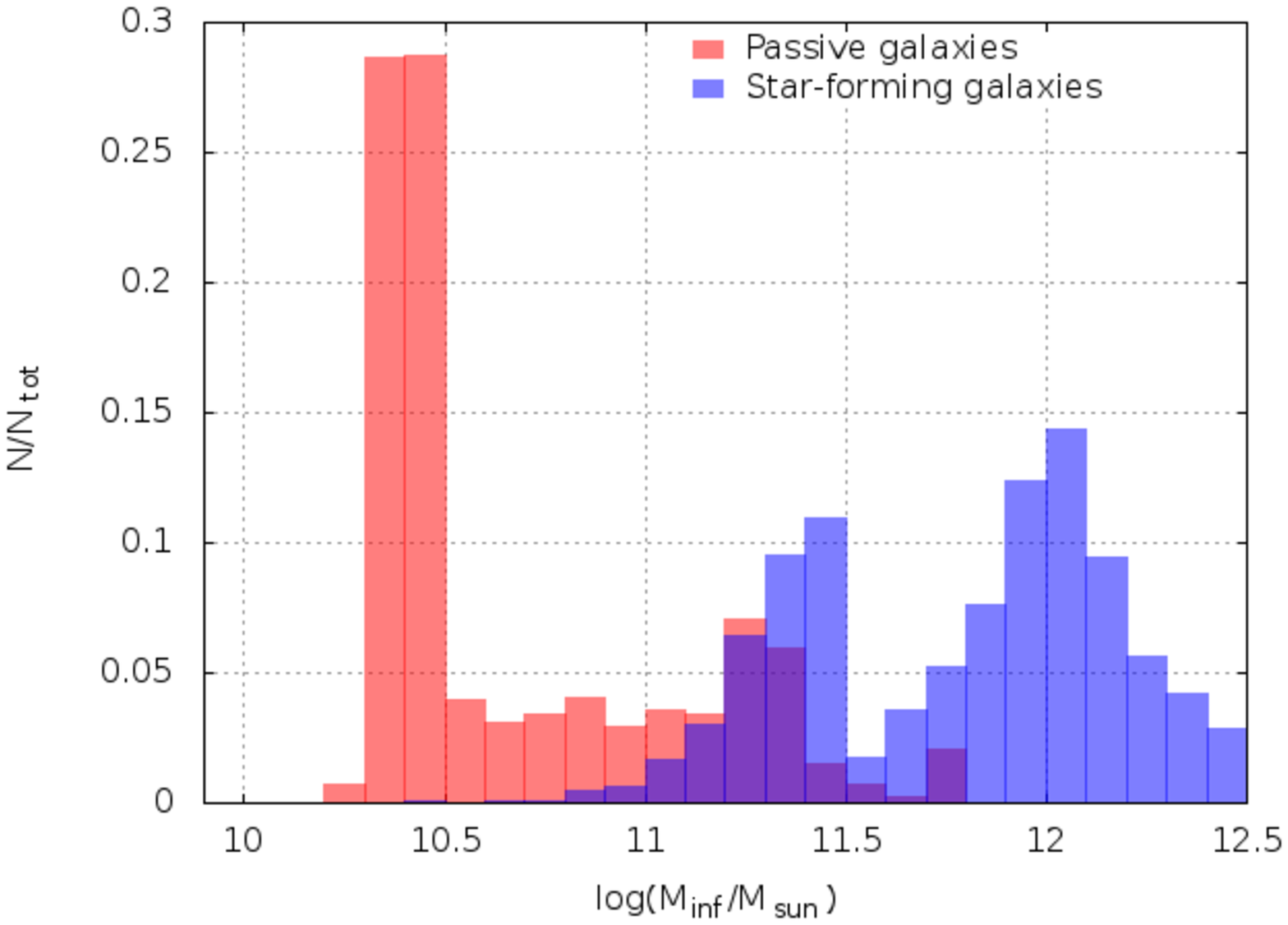} 
\includegraphics[scale=0.45]{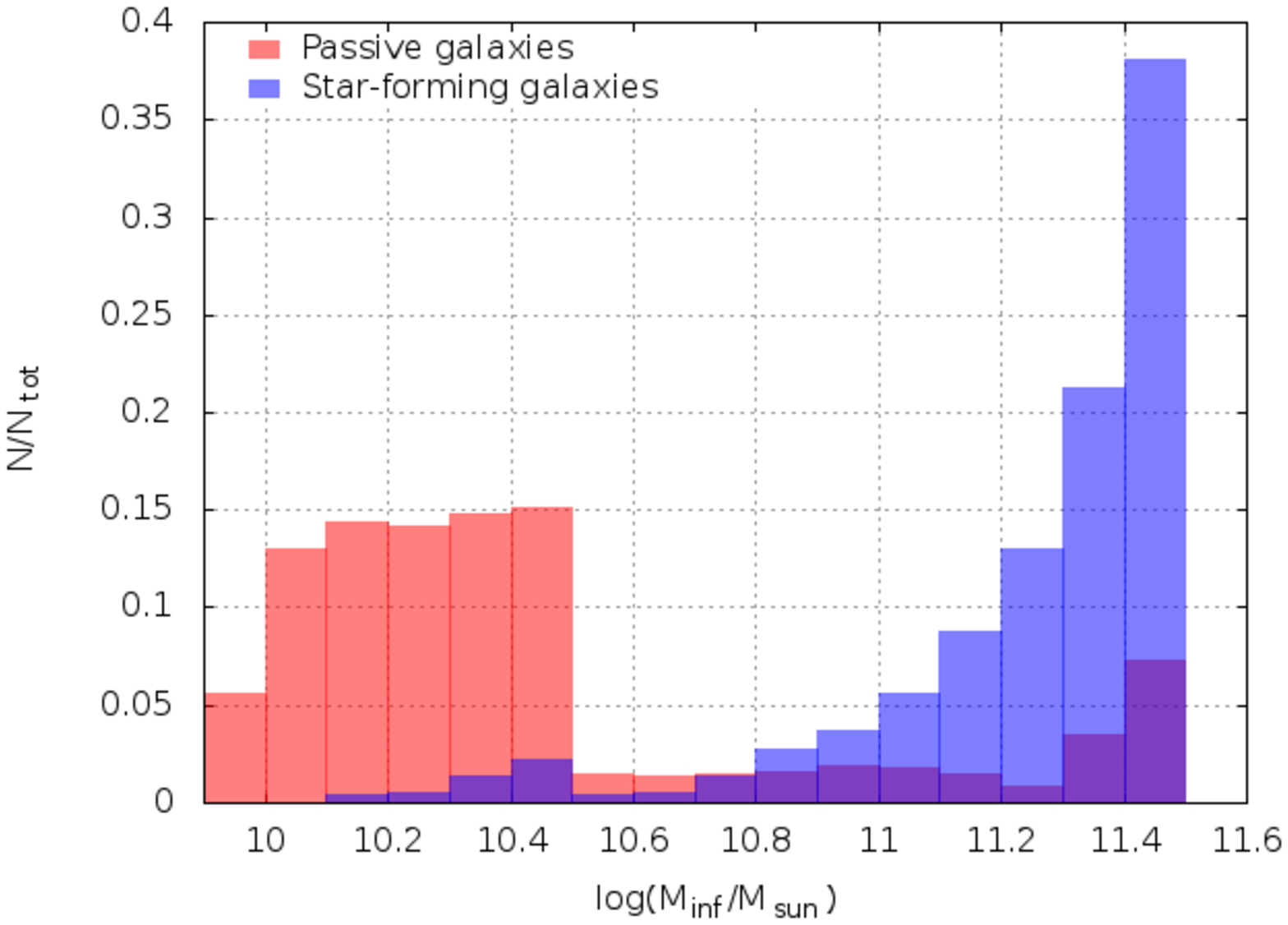} 
 \caption{{\it Upper panels}: As in Fig. \ref{TD} but here, the color
code indicates different infall mass $M_{inf}$ of the galaxies which
reproduce the MZ relations. {\it Lower panels}: The distribution of
the predicted passive galaxies (red histogram) and star-forming galaxies
(blue histogram) which reproduced the MZ relation in terms of infall
mass $M_{inf}$ of galaxies.  In the left panel we adopt a Chabrier (2003) IMF, whereas in the right one the Salpeter (1955) one;} \label{minf}
\end{figure*}

\subsection{Summary of the free parameters and constraints}

\label{sec:free_par_constraint}

The free parameters entering in the analytical solutions of our chemical evolution model are the following:
\begin{enumerate}
\item the infall timescale $\tau$;
\item the infall mass $M_\mathrm{inf}$;
\item the wind parameter $\lambda$.
\end{enumerate}
We created a set of chemical evolution models by varying these free parameters with a very fine resolution. 
 The assumed  infall timescale $\tau$ spans the range between 0.1 and 8 Gyr, with a resolution of the grid values  of $\Delta \tau$=0.05 Gyr. The  wind parameter  $\lambda$  is defined between 0 and 10,  with a
resolution  of $\Delta \lambda$=0.5. The infall masses are in the range between $10^{5.5}$ and 10$^{12.5}$ M$_{\odot}$.

 The SFE
is determined by means of Eq. (\ref{taugas}), where $M_{\star}$ is
replaced with $M_\mathrm{inf}$.  We cannot vary the SFE of the
galaxy according to its stellar mass, since the system of
Eq. (\ref{system1}) is solved by keeping the SFE constant.

The observational constraints that we assume to characterize the
star-forming and passive populations of galaxies in the SDSS sample
of \citet{peng2015} are the following:
\begin{enumerate}
\item Our models for passive and star-forming galaxies have average stellar metallicities defined such that \\
$|\log(Z_{\star,\mathrm{mod}})-\log (Z_{\star,\mathrm{obs}})|< 0.001 \, \mathrm{dex}$, where
$Z_{\star,\mathrm{mod}}$ are computed with Eq. (\ref{mass-weighted_Z})
and $Z_{\star,\mathrm{obs}}$ are the  stellar metallicities obtained with the fit reported in Fig. \ref{obs} for the  observed values
by \citet{peng2015}.
\item Our models for star-forming galaxies have  $|\mu_{\star,\mathrm{mod}}-\mu_{\star,\mathrm{obs}}|< 0.001$, 
where $\mu_{\star,\mathrm{obs}}$ are computed by means of Eq. (\ref{mustar}). 
\item Our models for passive galaxies must have $\mathrm{sSFR}_\mathrm{mod,passive} <2.29 \times 10^{-11}\,\mathrm{yr}^{-1}$, 
where $\mathrm{sSFR}_\mathrm{mod}$ is the predicted specific SFR.
\item Our models for star-forming galaxies must have $\mathrm{sSFR}_\mathrm{mod,star-forming} >2.29 \times 10^{-11}\,\mathrm{yr}^{-1}$. 
\end{enumerate}
\noindent
In this way, we are able to select and distinguish which 
chemical evolution model best represents the sample of \citet{peng2015}.  We
follow the chemical evolution of our entire galaxy sample with a fixed
time resolution $\Delta t = 0.0065\,\mathrm{Gyr}$.  The ages of all
star-forming and passive galaxies are defined as the galactic time when all
the observational constraints for a given population are fulfilled.

  As stressed before, we study the case of primordial
infall of gas (i.e., $Z_{inf}=0$). Although \citet{lehner2013} studied
the circumgalactic medium of galaxies and recently demonstrated that metal-enriched infalls occur, previous papers (\citealt{tosi1988,
matteucci2012}) have shown that an infall enriched with
metallicity $Z \leq$ 0.4 $Z_{\odot}$ does not produce differences in
the evolution of the solar neighborhood.  Moreover, assuming an infall
with a metallicity higher than 0.4 $Z_{\odot}$ requires very specific
situations. Recently, \citet{spitoni2016} considered the effects
of an enriched infall of gas with the same chemical abundances as the
matter ejected and/or stripped from dwarf satellites of the Milky Way
on the chemical evolution of the Galactic halo. We found that $\alpha$
elements are only slightly affected by such an enriched infall of gas.

\begin{figure*}
\centering
     \includegraphics[scale=.8]{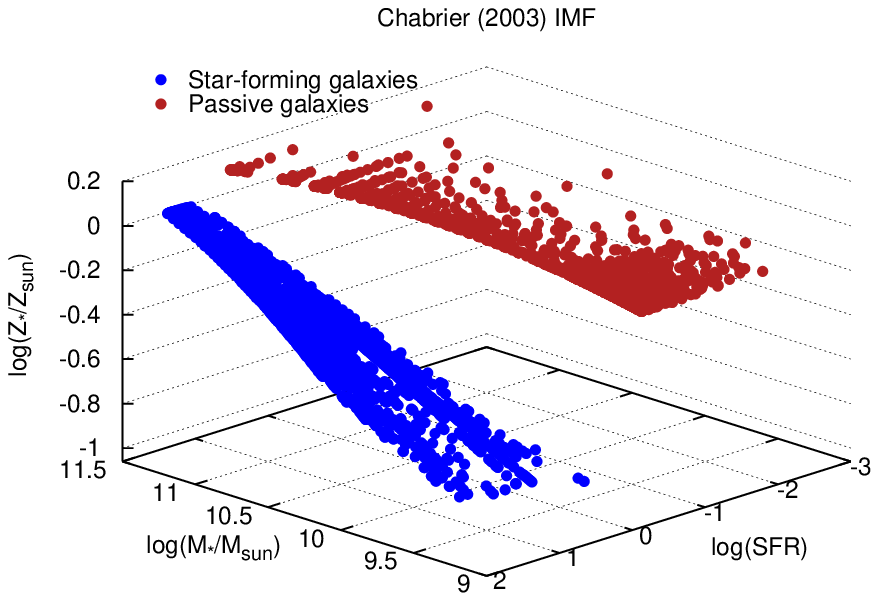}
\includegraphics[scale=.8]{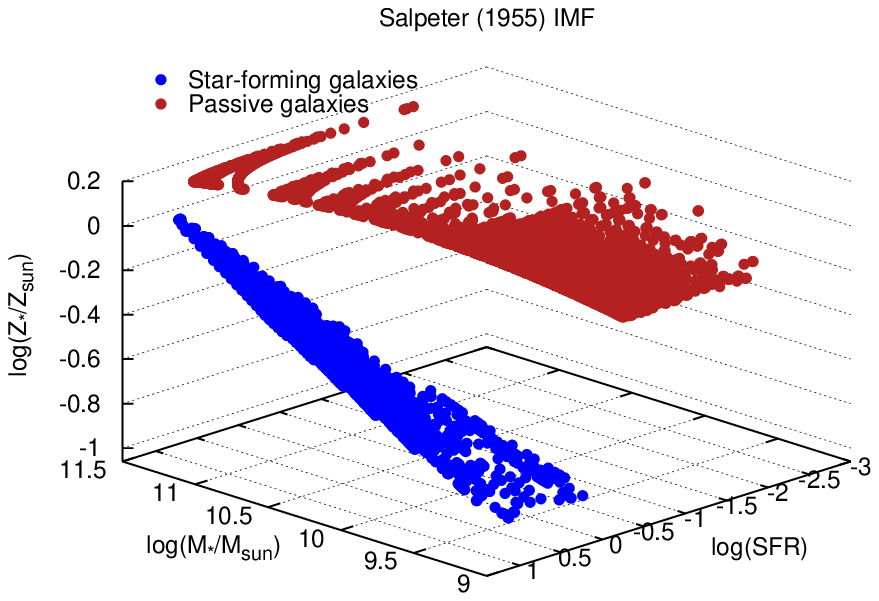}
\\
 \includegraphics[scale=.8]{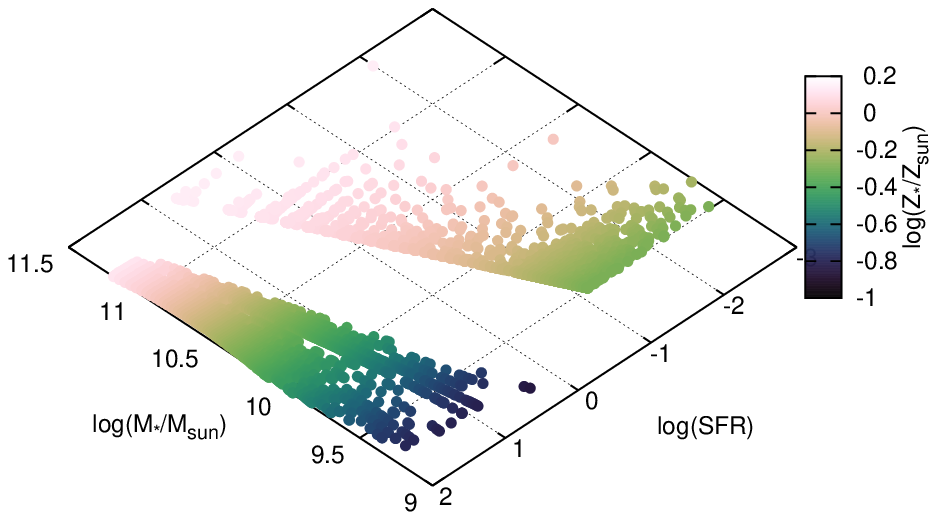}
 \includegraphics[scale=.8]{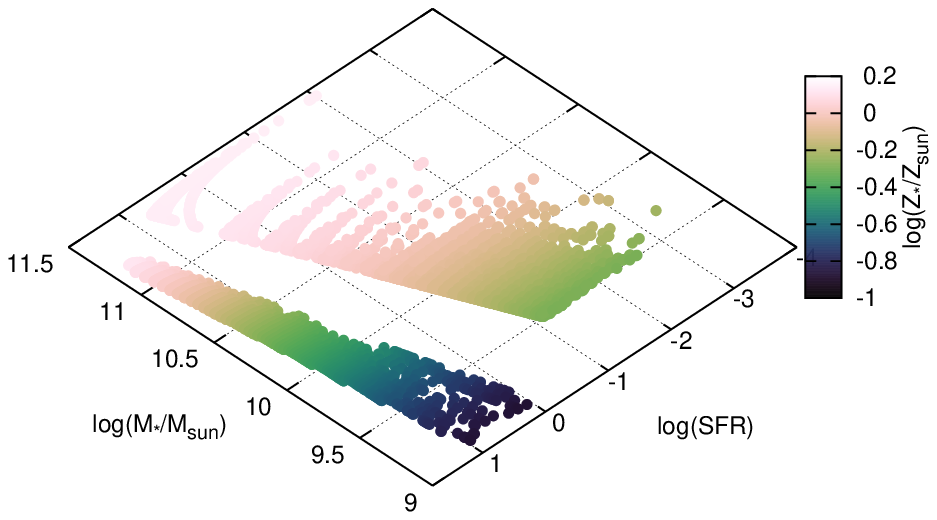}
\caption{{\it Upper panels:} Simulated passive and star-forming galaxies that are able to reproduce the MZ relations of Peng et al. (2015) in the 3D plot to test a more general relation between stellar $M_*$, average stellar metallicity $Z,$ and the SFR using a Chabrier (2003) (left panel) and a Salpeter IMF (right panel). {\it Lower panels:} The 2D projection in the $M_*$- SFR plane. }
 \label{FUNDA1}
\end{figure*}

\section{Results}

In this section we present our results concerning the characterization
of the local passive and star-forming galaxies of the SDSS sample
of \citet{peng2015}, by making use of the new analytical solutions
presented in Sect. \ref{sec:analytical_solutions} in the presence of an
exponential infall of gas.  We remark on the fact that we selected the
best chemical evolution models characterizing the SDSS sample
of \citet{peng2015} by imposing the set of constraints presented in
the previous Sect. \ref{sec:free_par_constraint} and by varying all
the free parameters simultaneously.

In Table \ref{table} we summarize the main results of our work.  We
report the range of values spanned by the main free parameters of our
models to reproduce the observed MZ relation of the local
star-forming and passive galaxies. For each parameter we also report the
value within which $75$ per cent of the galaxies are expected to
reside.  We show in the table our results for both
the \citet{salpeter1955} and the \citet{chabrier2003} IMFs.  A
fundamental quantity that we can predict is the age distribution of
the passive and star-forming galaxy SDSS population, corresponding to the
second row in the table.

 We assume
 that galaxies can form at different times, and  by age  we mean the galaxy evolutionary time, namely the difference between the time corresponding  to  redshift $z=\left<0.05\right>$  and the formation epoch.

To better visualize our results, in
Figs. \ref{TD}-\ref{minf} we show how local
star-forming and passive galaxies are expected to be distributed in the MZ
relation (upper panels) and in relative number (lower panels) for
different values of the infall timescale, age, wind parameter, and
infall mass, respectively.   In the lower panels of the figures we
indicate with N/N$_{tot}$ the ratio between the number of the computed
star-forming (or passive) galaxies in the considered bins of infall
timescale, age, wind parameter, and infall mass values over the total
number of the computed star-forming (passive) galaxies. The plots on
the left correspond to our results with a \citet{chabrier2003} IMF,
while on the right we show our results with a \citet{salpeter1955}
IMF.

 Figure \ref{TD} shows that our models for the local passive
 galaxies are characterized by shorter typical formation timescales
 than the star-forming galaxies. This is mainly due to the requirement of very
 low sSFRs for these galaxies at the present time. Therefore, they
  had the time to consume or remove most of their total infall gas
 mass through the star formation activity or galactic wind.  The short
 typical formation timescales also enhance the star formation
 activity and hence the metal production in the earliest epochs of the
 galaxy evolution.  According to the predictions of our models, local
 passive galaxies are currently undergoing the declining,
 fading phase of their SFR evolution.

 \par In our analytical model of chemical evolution, the IMF only
 enters in the calculation of the yield of metals per stellar
 generation  and  the returned mass fraction. Top-heavy IMFs
 determine higher yields of metals per stellar generation and hence a
 more effective chemical enrichment of the galaxy ISM, fixed all the
 other model parameters.  Therefore, passive galaxies with a
 top-heavy IMF that reach their relatively high observed stellar
 metallicity are on average characterized by longer formation timescales (a less intensive SFR at early times).  We find that when
 we assume the \citet{salpeter1955} IMF, the distribution of the formation
 timescales of passive galaxies is narrower than when
 we assume a \citet{chabrier2003} IMF, which contains a larger number
 of massive stars.  We find that almost $75$ per cent of all passive
 galaxies are expected to assemble on $\tau \le 1.8\,\mathrm{Gyr}$
 with a \citet{salpeter1955} IMF, while this time is $\tau \le2.4\,\mathrm{Gyr}$
 with a \citet{chabrier2003} IMF (see also Table \ref{table}).

\par In Fig. 4 we show the results of our models for the age distribution of 
local galaxies. Our main result is that passive galaxies are on
average much older than the star-forming galaxies. The passive galaxies need 
a longer  evolution to exhaust their total reservoir of gas, as described before, and to be observed as
quiescent objects at the present time.

\par In the case of star-forming galaxies 
 the
star-forming galaxies with higher mass generally show older ages and longer typical
formation timescales than star-forming
galaxies with lower masses.

  To reach their
higher observed stellar metallicities with respect to the less massive
galaxies, the more massive galaxies must be older.
On the other hand, longer formation timescales can ensure more
massive galaxies to be still observed as star forming at the present
time, since large amounts of gas may be still available for star
formation at later epochs.

\citet{peng2015} were able to also derive luminosity-weighted stellar 
ages for their galaxy sample. Their main finding is that passive
galaxies are on average about $4\,\mathrm{Gyr}$ older than the
star-forming galaxies, with this difference remaining almost constant when
considering different bins for the galaxy stellar mass.

   Considering the difference between the median age values of our
computed star-forming and passive galaxies, we find that our results
are in agreement with the data by \citet{peng2015}.  With
the \citet{chabrier2003} IMF, we find that the median stellar ages for
 star-forming and passive galaxy are $0.9\,\mathrm{Gyr}$ and
$5.1\,\mathrm{Gyr}$, respectively; therefore the $\Delta\, \mathrm{age}
=4.2 \,\mathrm{Gyr}$.  We find a larger difference between the median ages
when we adopt the \citet{salpeter1955} IMF; in this case, the median
stellar ages for  star-forming and passive galaxies are
$2.1\,\mathrm{Gyr}$ and $7.5\,\mathrm{Gyr}$, respectively; therefore\
$\Delta \, \mathrm{age}                       
=5.4 \,\mathrm{Gyr}$.

 Models of  galaxies with a
 top-heavy IMF are expected to need less time to reach their
 relatively high observed stellar metallicity because of the  chemical enrichment, which is driven by
a  larger number of massive stars; the minimum age we found for passive galaxies with the \citet{chabrier2003} IMF is 0.6 Gyr, much younger than the value we found using  the \citet{salpeter1955} IMF (1.8 Gyr).

In Fig. \ref{L} we present our results for the distribution of the
wind parameter, $\lambda$, associated with our best models for the
star-forming and passive galaxy populations of the SDSS sample
of \citet{peng2015}.  According to our results, the population of
galaxies that suffer more prominent wind episodes are the
star-forming galaxies.  To reproduce the observed average stellar
metallicity of star-forming galaxies at lower stellar mass, stronger
winds are needed.  On the other hand, the wind parameter, $\lambda$,
is predicted to decrease toward star-forming galactic
systems with higher stellar mass.  In summary, the observed MZ
relation can be well reproduced by assuming a variable mass-loading
factor, which increases when passing from more massive to less massive
galactic systems. This conclusion has been discussed by several
authors in the past (see, for example, \citealt{spitoni2010}, and
references therein).  This correlation between the mass-loading
factor, $\lambda$, and the galaxy stellar mass is also valid for
passive galaxies, although the galactic winds for these systems are
predicted to be much weaker than for the star-forming galaxies.

The IMF choice can also affect the typical values of the wind parameter  
that best reproduce the observed MZ relation. Our best models for low-mass 
and metal-poor star-forming galaxies with a \citet{salpeter1955} IMF require 
weaker galactic winds than similar models with a \citet{chabrier2003} IMF, 
which predict a faster and hence more effective chemical enrichment of the 
galaxy ISM. 
Table \ref{table} and Fig. \ref{L} show that the distribution of the wind parameters  
for star-forming galaxies with a \citet{salpeter1955} IMF span the range $0\la\lambda\la 5.8$,   and the 75\% of star-forming galaxies are expected to have $\lambda\la 2.25$, while the models of star-forming galaxies with a \citet{chabrier2003} IMF 
span the range $0.6\la\lambda\la 10.2$, and   and the 75\% of star-forming galaxies are expected to have $\lambda\la 5.3$.

 At variance with  \citet{peng2015}, our models fit the data when galactic winds occur during the entire galaxy evolution.

\begin{figure}
          \centering 
\includegraphics[scale=0.7]{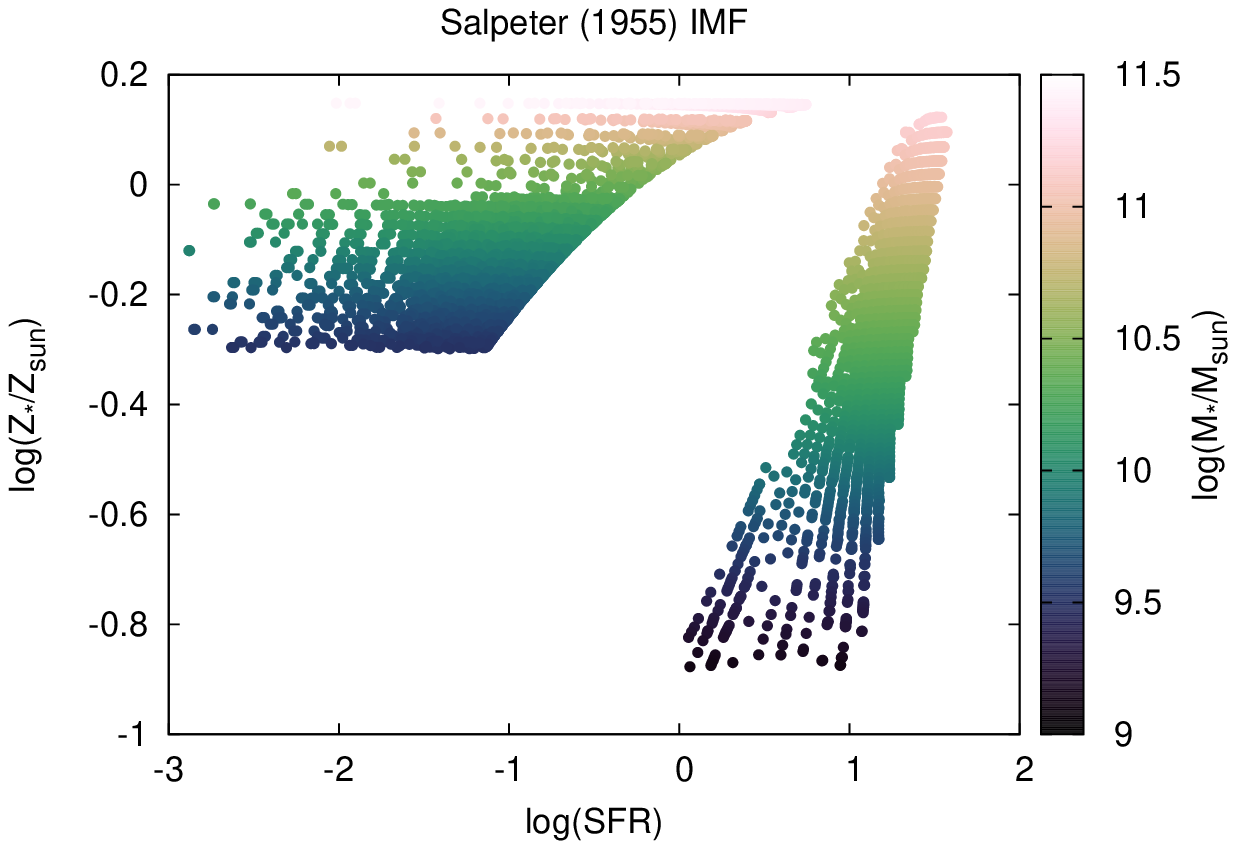}
\includegraphics[scale=0.7]{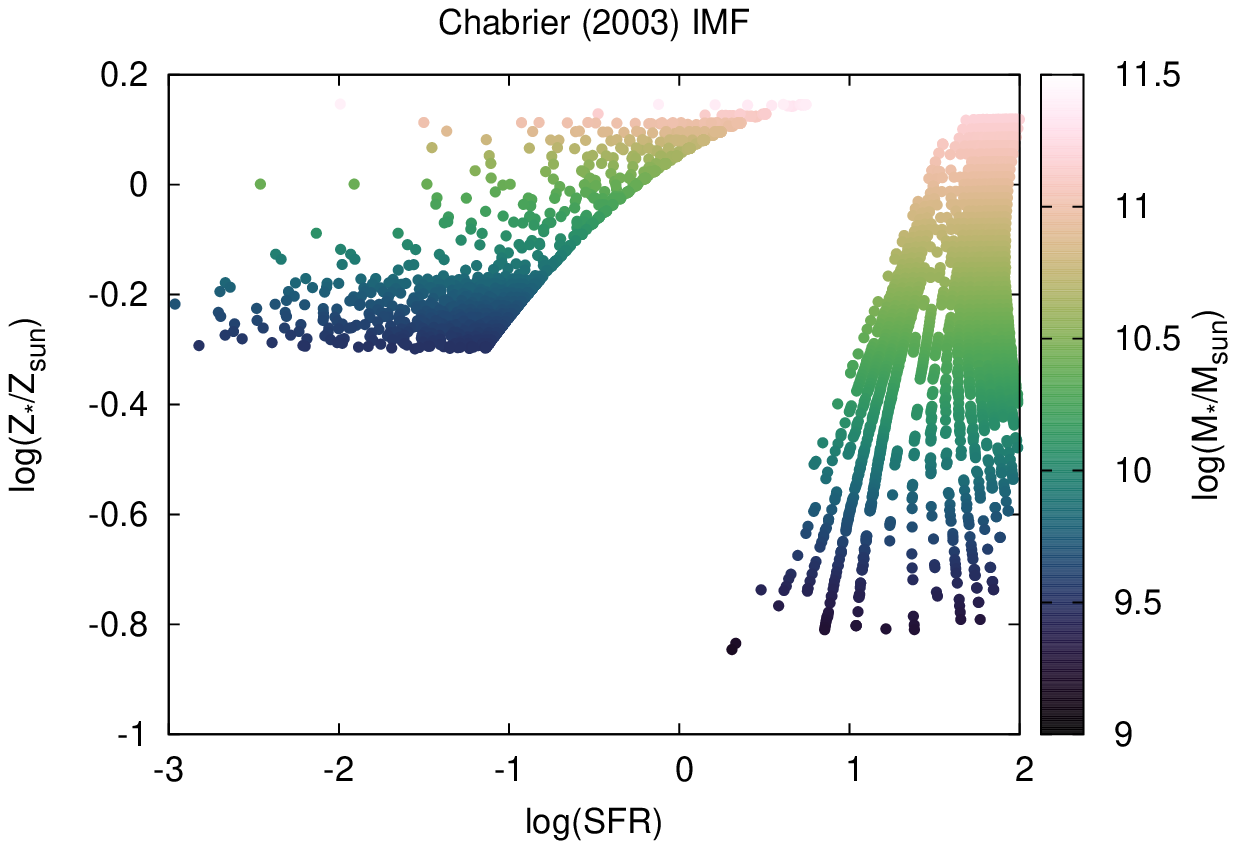}
 \caption{ Stellar metallicity as a function of the SFR  for the computed star-forming and passive galaxies. The color
code indicates the stellar mass. In the {\it upper panel} model results with a Salpeter (1955) IMF are drawn, and in the   {\it lower panel} a Chabrier (2003) IMF is adopted.} \label{FUNDA2}
\end{figure} 

In Fig. \ref{minf} we show our results for the distribution of the infall mass 
of our best models for the local population of star-forming and passive galaxies. 
We find that passive galaxies are characterized by
lower infall masses than the star-forming galaxies. 
Passive galaxies in the Local Universe must be characterized by very low 
sSFRs, which can be obtained in the framework of our analytical model with 
an exponential infall of gas by assuming low infall gas masses together 
with short typical timescales (as we discussed above for the passive population). 
In this way, nearly most of the infall gas mass is converted into stars 
at the later epochs of the galaxy evolution. 
Typical values of the infall mass of our best models for the passive and star-forming galaxies 
can be found in Table \ref{table}. 

 The results for the infall mass distribution of the local population of
 star-forming and passive galaxies are strongly affected by the
 adopted IMF.  Higher infall masses are required with a  \citet{chabrier2003} IMF.  We
 recall here that we assumed a primordial infall of gas. Because a
 top-heavy IMF leads to a more efficient chemical enrichment than
 a \citet{salpeter1955} IMF, a stronger dilution of metals and therefore higher infall masses are required to obtain the observed MZ relation.

\section{Predicted fundamental relation of the local star-forming and passive galaxies}

In this section, we analyze the correlations that local galaxies are predicted to show  
by our chemical evolution model in the 3D space defined by stellar mass, gas-phase 
metallicity, and SFR. \citet{mannucci2010}  showed that local galaxies 
lie on a tight surface in such a space and named the resulting correlation 
the ``fundamental relation'' , implying that it is valid at all redshifts. 

\par Since we do not possess gas-phase metallicities for the  passive galaxies in the  sample of \citet{peng2015}, 
we use the average galaxy stellar metallicity instead of the gas-phase
metallicity  to study the MZ relation.  The gas metallicity is
typically higher than
the stellar  one for evolved galactic systems.  Our results are shown
in Fig. \ref{FUNDA1}, where we show that local star-forming and
passive galaxies lie on a tight surface when considering their stellar
mass, $M_{\star}$, mass-weighted stellar metallicity,
$\left<Z_{\star}\right>$, and SFR. The projection of this 3D diagram
onto the $M_{\star}$-$Z_{\star}$ place gives rise to the MZ relation
that we discussed in the previous sections. The figure shows
that the two
populations of galaxies are separated by a discontinuity that is due
to the different metallicity distributions of star-forming and passive
objects.

\par The discontinuity is also visible in Fig. \ref{FUNDA2}, where we
show the projection of the predicted fundamental relation onto the
$Z_{\star}$-$\mathrm{SFR}$ plane.  In this figure, the color-coding
corresponds to the galaxy stellar mass.  The local galaxy populations
are predicted to show a similar trend as the one presented
by \citet[see the right panel of their Fig. 1]{mannucci2010}, where
higher stellar masses are found for galaxies with higher SFRs and
metallicities.

\section{Conclusions}

We have developed a new analytical model of chemical
evolution, in which an exponential infall of gas and galactic winds
are assumed.  We applied this model to reproduce the observed MZ
relation for local SDSS galaxies (\citealt{peng2015}); in particular,
we characterized the populations of gas-rich and star-forming and
passive (gas-poor and quiescent) galaxies, by showing how their ages,
formation timescales, mass-loading factors and infall masses must
relate to each other. Finally, we analyzed the fundamental
relation for these local galaxies in the 3D space, defined by stellar
mass, average stellar metallicity, and SFR.  Our main conclusions can
be summarized as follows:

\begin{enumerate}

\item 
 We assume that
all galaxies form by gas accretion with an exponential law.
 We find that passive galaxies are characterized by shorter typical formation timescales and are older objects than the star-forming galaxies. 

\item Galactic winds in star-forming galaxies are found  on average to be stronger than in passive galaxies. 
The intensity of the galactic winds depends on the adopted IMF, with
the higher values of the mass-loading factor corresponding to
top-heavy IMFs. This is a consequence of the fact that top-heavy
IMFs lead to a more  efficient metal enrichment of the galaxy ISM than
bottom-heavy IMFs.

\item The observed MZ relation of \citet{peng2015} can be reproduced by our models without invoking any strangulation effects. 
This is because the assumption of an exponential infall
of gas, also coupled to galactic winds, naturally  reduces the gas
accretion after the assumed infall timescale, hence mimics the effect
of strangulation.  Nevertheless, we conclude that strangulation is not
the main physical mechanism driving the transition of galaxies toward
the passive evolution.

\item We have shown that our models for star-forming and passive galaxies  imply that they obey the so-called fundamental relation of 
\citet{mannucci2010}, which a is more general relation between stellar mass, metallicity, and SFR. The fundamental relation of \citet{mannucci2010} adopts the
gas-phase metallicity,  and we find that  it is still valid  when adopting the
average galaxy stellar metallicity  ($M_{\star}$ - SFR - $Z_{\star}$). 
\end{enumerate}

\section*{Acknowledgments}
We thank the anonymous referee for the suggestions that
improved the paper.
 The work was supported by PRIN MIUR 2010-2011, project ``The Chemical
and dynamical Evolution of the Milky Way and Local Group Galaxies'',
prot. 2010LY5N2T.


\begin{thebibliography}{99}

\bibitem[\protect\citeauthoryear{Boissier \& Prantzos}{1999}]{boissier1999}
Boissier, S., \& Prantzos, N., 1999, MNRAS, 307, 857 

\bibitem[\protect\citeauthoryear{Boselli et al.}{2014}]{boselli2014}
 Boselli, A., Cortese, L., Boquien, M., et al., 2014, A\&A, 564, A66 
 
\bibitem[\protect\citeauthoryear{Boselli et al.}{2002}]{boselli2002}
Boselli, A., Lequeux, J.,  Gavazzi, G. 2002, A\&A, 384, 33

\bibitem[\protect\citeauthoryear{Brusadin et al.}{2013}]{brusadin2013}
Brusadin G., Matteucci F., Romano D., 2013, A\&A, 554, A135

\bibitem[\protect\citeauthoryear{Chabrier}{2003}]{chabrier2003}
Chabrier, G., 2003, PASP, 115, 763


\bibitem[\protect\citeauthoryear{Chiappini et al.}{1997}]{chiappini1997}
Chiappini C., Matteucci F., Gratton R., 1997, ApJ, 477, 765

\bibitem[\protect\citeauthoryear{Chiosi}{1980}]{chiosi1980}
Chiosi, C.,  1980, A\&A, 83, 206 

\bibitem[\protect\citeauthoryear{Clayton}{1988}]{clayton1988}
Clayton, D. D., 1988, MNRAS, 234, 1
\bibitem[\protect\citeauthoryear{Clayton \& Pantelaki}{1986}]{clayton1986}
 Clayton, D. D.,  Pantelaki, I.,  1986, ApJ 307, 441 
\bibitem[\protect\citeauthoryear{Clayton \& Pantelaki}{1993}]{clayton1993}
 Clayton, D. D., Pantelaki, I., 1993, PhR 227, 293)
\bibitem[\protect\citeauthoryear{Colavitti et al.}{2008}]{colavitti2008}
Colavitti, E., Matteucci, F.,  Murante, G., 2008, A\&A, 483, 401 


\bibitem[\protect\citeauthoryear{Edmunds}{1990}]{edmunds1990}
Edmunds, M. G., 1990, MNRAS, 246, 678




\bibitem[\protect\citeauthoryear{Edmunds \& Greenhow}{1995}]{edmunds1995} Edmunds M.~G., Greenhow R.~M., 1995, MNRAS, 272, 241 

\bibitem[\protect\citeauthoryear{Erb}{2008}]{erb2008}
Erb, D.~K.\ 2008, ApJ, 674, 151-156 


\bibitem[\protect\citeauthoryear{Fossati et al.}{2015}]{fossati2015}
Fossati, M., Wilman, D. J., Fontanot, F., et al.,  2015, MNRAS, 446, 2582 

\bibitem[\protect\citeauthoryear{Franx et al.}{2008}]{franx2008} Franx M., van Dokkum P.~G., F{\"o}rster Schreiber N.~M., et al., 2008, ApJ, 688, 770-788 


\bibitem[\protect\citeauthoryear{Hartwick}{1976}]{hartwick1976} Hartwick F.~D.~A., 1976, ApJ, 209, 418 



\bibitem[\protect\citeauthoryear{Kewley \& Ellison}{2008}]{kewley2008}
Kewley, L. J.,  Ellison, S. L. 2008, ApJ, 681, 1183
\bibitem[\protect\citeauthoryear{Kroupa}{2001}]{kroupa2001}
Kroupa, P., 2001, MNRAS, 322, 231 
\bibitem[\protect\citeauthoryear{Kudritzki et al.}{2015}]{kudritzki2015}
Kudritzki, R.-P., Ho, I.-T., Schruba, A., et al., 2015, MNRAS, 450, 342 
\bibitem[\protect\citeauthoryear{Lacey \& Fall}{1985}]{lacey1985}
Lacey, C. G.,  Fall, M., 1985, ApJ, 290, 154

\bibitem[\protect\citeauthoryear{Lanfranchi et al.}{2008}]{lanfranchi2008}
Lanfranchi G. A., Matteucci F., Cescutti G., 2008, A\&A, 481, 635

\bibitem[\protect\citeauthoryear{Lehner et al.}{2013}]{lehner2013}
 Lehner, N., Howk, J.~C., Tripp, T.~M., et al., 2013, ApJ, 770, 138 



\bibitem[\protect\citeauthoryear{Mannucci et al.}{2010}]{mannucci2010}
Mannucci, F., Cresci, G., Maiolino, R., Marconi, A.,  Gnerucci, A., 2010, MNRAS, 408, 2115 

\bibitem[\protect\citeauthoryear{Martinelli}{1998}]{martinelli1998} Martinelli A., 1998, A\&A, 335, 847 
\bibitem[\protect\citeauthoryear{Matteucci \& Chiosi}{1983}]{matteucci1983}
Matteucci, F.,  Chiosi, C., 1983, A\&A, 123, 121 
\bibitem[\protect\citeauthoryear{Matteucci}{2001}]{matteucci2001}
Matteucci, F. 2001, The Chemical Evolution of the Galaxy, ASSL, Kluwer
Academic Publisher

\bibitem[\protect\citeauthoryear{Matteucci}{2012}]{matteucci2012}
Matteucci F., 2012, Chemical Evolution of Galaxies. Springer-Verlag, Berlin

\bibitem[\protect\citeauthoryear{Micali et al.}{2013}]{micali2013}
Micali A., Matteucci F., Romano D., 2013, MNRAS, 436, 1648


\bibitem[\protect\citeauthoryear{Pagel \& Patchett}{1975}]{pagel1975}
Pagel, B. E. J.,  Patchett, B. E. 1975, MNRAS, 172, 13

\bibitem[\protect\citeauthoryear{Pagel}{1997}]{pagel1997}
Pagel B. E. J., 1997, in Bernard E. J. P., eds, Nucleosynthesis and Chemical Evolution of Galaxies, ISBN: 0521550610. Cambridge Univ. Press,
Cambridge, p. 392


\bibitem[\protect\citeauthoryear{Peng et al.}{2015}]{peng2015}
Peng, Y., Maiolino, R.,  Cochrane, R., 2015, Nature, 521, 192 

\bibitem[\protect\citeauthoryear{Peng \& Maiolino}{2014}]{peng2014}
Peng, Y., Maiolino, R., 2014, MNRAS, 443, 3643


\bibitem[\protect\citeauthoryear{Pezzulli \& Fraternali}{2015}]{pezzulli2015}
Pezzulli, G., Fraternali, F.,  2015, MNRAS, 455, 2308

\bibitem[\protect\citeauthoryear{Portinari \& Chiosi}{2000}]{portinari2000}
Portinari, L.,  Chiosi, C. 2000, A\&A, 355, 929
\bibitem[\protect\citeauthoryear{Recchi \& Kroupa}{2015}]{recchi2015} 
Recchi, S., Kroupa, P, MNRAS, 2015, 446, 4168 



\bibitem[\protect\citeauthoryear{Recchi et al.}{2008}]{recchi2008} 
Recchi, S., Spitoni, E., Matteucci, F., Lanfranchi, G. A., 2008, A\&A, 489, 555



\bibitem[\protect\citeauthoryear{Recchi \& Hensler}{2013}]{recchi2013}
Recchi, S., \& Hensler, G., 2013, A\&A, 551, A41 

\bibitem[\protect\citeauthoryear{Romano et al.}{2010}]{romano2010} 
Romano D., Karakas A. I., Tosi M., Matteucci F., 2010, A\&A, 522, A32


\bibitem[\protect\citeauthoryear{Salpeter}{1955}]{salpeter1955}
Salpeter, E. E. 1955, ApJ, 121, 161

\bibitem[\protect\citeauthoryear{Sandage}{1986}]{sandage1986}
Sandage A., 1986, A\&A, 161, 89

\bibitem[\protect\citeauthoryear{Schmidt}{1963}]{schmidt1963}
Schmidt, M. 1963, ApJ, 137, 758

\bibitem[\protect\citeauthoryear{Schmidt}{1959}]{schmidt1959}
Schmidt, M. 1959, ApJ, 129, 243

\bibitem[\protect\citeauthoryear{Searle \& Sargent}{1972}]{searle1972}
Searle, L.,  Sargent, W. L. W. 1972, ApJ, 173, 25

\bibitem[\protect\citeauthoryear{Spergel et al.}{2007}]{spergel2007}
Spergel, D. N., Bean, R., Dor, O., et al. 2007, ApJS, 170, 377
\bibitem[\protect\citeauthoryear{Spitoni}{2015}]{spitoni2015}
Spitoni, E., 2015, MNRAS, 451, 1090


\bibitem[\protect\citeauthoryear{Spitoni et al.}{2010}]{spitoni2010}
Spitoni E., Calura F., Matteucci F., Recchi S., 2010, A\&A, 514, A73

\bibitem[\protect\citeauthoryear{Spitoni \& Matteucci}{2011}]{spitoni2011}
Spitoni E., Matteucci F., 2011, A\&A, 531, A72
\bibitem[\protect\citeauthoryear{Spitoni et al.}{2016}]{spitoni2016}
 Spitoni, E., Vincenzo, F., Matteucci, F.,  Romano, D., 2016, MNRAS, 458, 2541 

\bibitem[\protect\citeauthoryear{Springel}{2005}]{springel2005}
Springel, V., 2005, MNRAS, 364, 1105
\bibitem[\protect\citeauthoryear{Tinsley}{1974}]{tinsley1974} Tinsley B.~M., 1974, ApJ, 192, 629 

\bibitem[\protect\citeauthoryear{Tinsley}{1980}]{tinsley1980} Tinsley B.~M., 1980, Fundamentals of Cosmic Physics, 5, 287 

\bibitem[\protect\citeauthoryear{Tosi}{1988}]{tosi1988}
Tosi, M., 1988, A\&A, 197, 33

\bibitem[\protect\citeauthoryear{Twarog}{1980}]{twarog1980}
Twarog B. A., 1980, ApJ, 242, 242

\bibitem[\protect\citeauthoryear{Vincenzo et al.}{2016}]{vincenzo2016}
Vincenzo, F., Matteucci, F., Belfiore, F.,  Maiolino, R., 2016, MNRAS, 455, 4183 
\bibitem[\protect\citeauthoryear{Vincenzo et al.}{2014}]{vincenzo2014}
Vincenzo F.,Matteucci F., Vattakunnel S., Lanfranchi G. A., 2014, MNRAS,
441, 2815

\bibitem[\protect\citeauthoryear{Weinberg et al.}{2016}]{weinberg2016}
Weinberg, D.H., Andrews, B.H.,  Freudenburg, J. 2016, arXiv:1604.07435 



\bibitem[\protect\citeauthoryear{Zibetti et al.}{2009}]{zibetti2009}
Zibetti, S., Charlot, S., Rix, F. 2009, MNRAS, 400, 1181

\end{thebibliography}
\end{document}